\documentclass[acmsmall]{acmart}

\AtBeginDocument{%
  \providecommand\BibTeX{{%
    \normalfont B\kern-0.5em{\scshape i\kern-0.25em b}\kern-0.8em\TeX}}}

\setcopyright{rightsretained}
\acmJournal{PACMHCI}
\acmYear{2024} \acmVolume{8} \acmNumber{CSCW1} \acmArticle{54} \acmMonth{4}\acmDOI{10.1145/3652867}

\received{January 2023}
\received[revised]{July 2023}
\received[accepted]{November 2023}

\renewenvironment{quote}{%
  \list{}{%
    \leftmargin0cm   
    \rightmargin\leftmargin
  }
  \item\relax
}
{\endlist}

\definecolor{RoyalBlue}{cmyk}{1, 0.50, 0, 0}

\usepackage{xspace}
\usepackage[breakable, theorems, skins]{tcolorbox}
\tcbset{enhanced}

\DeclareRobustCommand{\mybox}[2][RoyalBlue!20]{%
\begin{tcolorbox}[   
        breakable,
        left=0pt,
        right=0pt,
        top=0pt,
        bottom=0pt,
        colback=#1,
        colframe=#1,
        width=\dimexpr\columnwidth\relax, 
        enlarge left by=0mm,
        boxsep=5pt,
        ]
        #2
\end{tcolorbox}
}

\newcounter{findingsCounter}
\newcommand\finding[1]{\mybox{\textbf{Finding~\stepcounter{findingsCounter}\arabic{findingsCounter}:}\xspace#1}}

\usepackage{xcolor}
\newcommand{\revisions}[1]{{\color{blue}#1}}

\usepackage{ulem}
\usepackage{subfig}
\usepackage{enumitem}
\usepackage{booktabs}
\usepackage{multirow}
\usepackage{graphicx}
\usepackage{longtable}

\makeatletter
\gdef\@copyrightpermission{
  \begin{minipage}{0.2\columnwidth}
   \href{https://creativecommons.org/licenses/by/4.0/}{\includegraphics[width=0.90\textwidth]{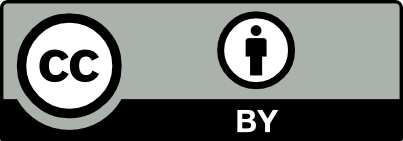}}
  \end{minipage}\hfill
  \begin{minipage}{0.8\columnwidth}
   \href{https://creativecommons.org/licenses/by/4.0/}{This work is licensed under a Creative Commons Attribution International 4.0 License.}
  \end{minipage}
  \vspace{5pt}
}
\makeatother
\begin{document}

\title{Exploring the Perspectives of Social VR-Aware Non-Parent Adults and Parents on Children's Use of Social Virtual Reality}


\author{Cristina Fiani}
\email{c.fiani.1@research.gla.ac.uk}
\orcid{0000-0002-7119-2383}
\affiliation{%
  \institution{University of Glasgow}
  \country{UK}
}
\author{Pejman Saeghe}
\email{Pejman.Saehghe@strath.ac.uk}
\orcid{0000-0001-6602-3123}
\affiliation{%
  \institution{University of Glasgow,}
  \country{UK}
}
\authornote{Helped with proof reading and road testing arguments for the qualitative analysis}
 
\author{Mark McGill}
\email{Mark.McGill@glasgow.ac.uk}
\orcid{0000-0002-8333-5687}
\affiliation{%
  \institution{University of Glasgow}
  \country{UK}
}
\author{Mohamed Khamis}
\email{Mohamed.Khamis@glasgow.ac.uk}
\orcid{0000-0001-7051-5200}
\affiliation{%
  \institution{University of Glasgow}
  \country{UK}
}


\renewcommand{\shortauthors}{Cristina Fiani, Pejman Saeghe, Mark McGill \& Mohamed Khamis}
\renewcommand{\shorttitle}{Exploring Perspectives of Social VR-Aware Parents and \\ Non-Parent Adults on Children's Use of Social Virtual Reality}

\begin{abstract}
Social Virtual Reality (VR), where people meet in virtual spaces via 3D avatars, is used by children and adults alike. Children experience new forms of harassment in social VR where it is often inaccessible to parental oversight. To date, there is limited understanding of how parents and non-parent adults within the child social VR ecosystem perceive the appropriateness of social VR for different age groups and the measures in place to safeguard children. We present results of a mixed-methods questionnaire (N=149 adults, including 79 parents) focusing on encounters with children in social VR and perspectives towards children's use of social VR. We draw novel insights on the frequency of social VR use by children under 13 and current use of, and future aspirations for, child protection interventions. Compared to non-parent adults, parents familiar with social VR propose lower minimum ages and are more likely to allow social VR without supervision. Adult users experience immaturity from children in social VR, while children face abuse, encounter age-inappropriate behaviours and self-disclose to adults. We present directions to enhance the safety of social VR through pre-planned controls, real-time oversight, post-event insight and the need for evidence-based guidelines to support parents and platforms around age-appropriate interventions.
\end{abstract}



\maketitle

\section{Introduction}
Immersive social virtual reality (VR) with VR headsets allows users to connect in novel ways, transforming social interactions from 2D web pages (social media e.g., Instagram) to 3D virtual spaces. 
As opposed to 2D social media in which users interact behind screens, in social VR, users interact via an embodied avatar synchronously in 3D immersive virtual environments, increasing the illusion of ``being there'' \cite{10.1145/3197391.3205451}. It has the potential to mimic true face-to-face interactions \cite{Maloney2020a}. Prior work has shown that young children respond differently to VR compared to traditional media: VR elicits differential cognitive and social responses compared to less immersive technology \cite{bailey2019, Miehlbradt2021}. 
Social VR is a disruptive technology that is currently growing. The past couple of years have seen VR reach the consumer market and grow significantly, especially with the pandemic accelerating adoption \cite{BARREDAANGELES2022, Gaggioli2018, BALL2021}. One of the mainstream social VR platforms, VRChat has an estimated number of 7.2 million players in total \cite{statsvrchat} and a peak of 127,919 players simultaneously \cite{mmostats}. Social VR, initially designed for adults, has attracted teenagers and younger children \cite{Maloney2021}. They have been drawn to social VR because of its engaging and immersive activities, allowing them to connect with friends beyond just playing games, especially since the COVID-19 pandemic \cite{Maloney2021}. 
On the downside, there has been an increase in harassment, bullying and new forms of harm in social VR \cite{10.1145/3359202, tseng2022chi, Maloney2021, Maloney2020b}. 
Children and adults have been reporting experiences of harassment, from name calling to physical stalking \cite{Maloney2021, Maloney2020b}.
Existing features such as blocking, personal space bubble, muting, reporting players \cite{AltspaceVRsafety, RecRoomSafety} or trust systems to keep users safe from nuisance users \cite{VRChatTrust} were found to be ineffective at times and can even be misused \cite{Maloney2020b}. 
As social VR is used via head-worn devices that completely occlude reality, and do not support bystander awareness by default \cite{joseph2021}, children's activity becomes more inaccessible to parental oversight. It becomes difficult to be aware of the misuses and abuses in social VR, denying parents the ability to observe, supervise and intervene. 
A growing number of parents are navigating a new frontier in technology. They may lack awareness and often experience \cite{GREENFIELD2004751}, and attempt to manage the impact of disruptive technology on the parent-child relationship \cite{Hartikainen2016, Hiniker2016, Blackwell2016}. Parents are therefore presented with the challenge of keeping up with their children's use and adapting their parenting strategies. These challenges have been tackled with regard to mobile phones and social media in particular \cite{Blackwell2016, Danet2020, Radesky2016}. Results also showed that improved parent-child relationships were correlated with parent involvement, mediation, co-viewing and open communication \cite{Blackwell2016}. However, the new and emergent parent-child tensions are yet to be researched in social VR.

While prior work studied why teenagers use social VR \cite{Maloney2021} and how children interact with other children and with adults in social VR \cite{Maloney2020a}, little is known about the perception of age appropriateness of Social VR, the need for supervision, and appropriate child safeguarding practices. 
In this paper, we investigate the following research questions:

\begin{itemize}[noitemsep, topsep=0pt]
    \item[\textbf{RQ1)}] How often do children use social VR based on parental estimates and what ages are perceived appropriate for its use?
    \item[\textbf{RQ2)}] What negative experiences and perceptions do parents and non-parent adults have of social VR involving minors? 
    \item[\textbf{RQ3)}] What kind of interventions or oversight do parents and non-parent adults deem necessary to protect children in social VR, and do parents note particular interventions that are more or less appropriate for differing age groups?
\end{itemize}

Addressing these questions will allow a better understanding of non-parent adults' and parents' concerns and children's actual experiences, which paves the way to building safeguarding tools and guidelines for improving children's safety in social VR. 

This paper presents empirical results of a mixed-methods questionnaire (149 responses) in which parents (N=79) and non-parent adults (N=70) reflected on their experiences with children in social VR. Both perspectives are important as they contribute to a new understanding of the experiences and issues of younger users in social VR. Parents are the legal guardians of children who may have a different view for their own children’s social VR usage. Non-parent adults that use social VR are likely to have relevant lived encounters with children, holding insights that parents may not necessarily have.
By examining these experiences through the lenses of the main adult stakeholders within the child social VR ecosystem, we gain invaluable insights into the dynamics and considerations that shape the child's social VR interactions. This includes encounters with children, as well as experiences of one's children in social VR. 
More specifically, our participants reported on notable experiences they have had in social VR that involved children, the estimated age of the children they encountered in said experiences, and the participant's feelings and actions were taken after the experience. 
Parents shared the estimated frequency of their children's usage of social VR, and current safeguarding practices and parental control tools they use to protect their children. 
Parents and non-parent adults also reflected on which ages, and on what interventions  they perceive to be appropriate for social VR. 

\subsection{Contribution Statement}

This paper makes a number of contributions to HCI and CSCW. We extend and enhance the knowledge regarding the evolving and emerging challenging interactions that occur between adults and children in social VR, as such an understanding underpins any attempt to safeguard both parties. 
Firstly, regarding the lived experiences of adults and children in social VR, existing studies, although few, have focused on the primary users' and teenagers’ perspectives. Our study offers a broader insight into social VR experiences by capturing perspectives of ecosystem adult stakeholders of youth social VR users including parents and non-parent adults encountering children in social VR.
Secondly, as opposed to prior studies, we also seek insight into how parents would choose to intervene and supervise their children on these new platforms, guiding research into age-appropriate safety-enhancing technology and guidelines around usage. These contributions are important as parents are the legal guardians of children, and will desire to be involved in the moderation and safeguarding of their child's online VR experiences. 
Taken together, our results expand and reveal new insights into the extent to which minors use social VR; how in our sample parenthood, familiarity and supervision influence perceptions of children's use; and how parents would choose to moderate otherwise limit their child's usage of social VR across their childhood (in particular findings \hyperref[sec:finding1]{1}, \hyperref[sec:finding2]{2},
\hyperref[sec:finding6]{6}, \hyperref[sec:finding7]{7}, \hyperref[sec:finding8]{8}).
Negative experiences with minors in social VR, perceptions and concerns described by parents and non-parent adults confirm results in prior work (findings \hyperref[sec:finding3]{3}, \hyperref[sec:finding4]{4}) and add knowledge regarding parental views to the field of social VR (finding \hyperref[sec:finding5]{5}). Based on our findings, we discuss how to make social VR a safer space, and present future work directions. 

\subsection{Terminology}

In this paper, we consider \textit{children} as minors aged under 18. Respondents to our questionnaire include \textit{parents} and \textit{non-parent adults}, which we address both in combination and separately throughout the paper. Where we discuss \textit{participants}, we always refer to the combination of both parents and non-parent adults (i.e. the full dataset of the questionnaire). We also refer to \textit{adult social VR users} as the general population of social VR users.

\section{Related Work}

Our work builds upon prior work on 1) youth in VR and social VR, 2) interactions between adults and minors in social VR, and 3) technology tensions among parents and children.

\subsection{Increase of Minors in VR and Social VR}

Over the past decade, research on children's use of VR has mainly focused on the effectiveness of VR in educational or medical settings for children \cite{Zhang2022, Didehbani2016, Mado2022}. In particular, VR has been used as an intervention tool for children with autism spectrum disorder \cite{Didehbani2016}, and for rehabilitation and therapy in children with acquired brain injuries \cite{Zhang2022}. With the pandemic, the need for remote instruction tools increased and VR has allowed immersive learning, shown to be more effective than other traditional tools \cite{Mado2022}. The latter study explored the innovative opportunities and challenges associated with using VR for children’s remote education via an online survey (N = 311) \cite{Mado2022}. They collected data from the parents and legal guardians of children who used VR at home during the first months of the global pandemic. The study showed that 71.5\% of participants' children increased their VR usage. 
Maloney and his colleagues explored minors' social VR use, not only presenting opportunities for this disruptive technology but also raising challenges about children's social VR use \cite{Maloney2020a, Maloney2020b, Maloney2021}. Social VR, initially designed for adults, has become widely used by teenagers and children \cite{Maloney2021} even though most platforms' recommendations require users to be at least 13 years. Social VR allows a wide range of activities and connections, and children's engagement especially grew during the global pandemic where they could keep in touch with friends. In particular, they observed the attributes that contribute to what attracts or dissuades teenagers from using social VR \cite{Maloney2021}. As part of the pitfalls dissuading the use of social VR, teenagers mentioned the occurrence of harassment and bullying in social VR. 
While this showed reasons why teenagers and young children may use social VR, to date it is not clear how frequently children use social VR.

\subsection{Adult and Minors Interactions in Social VR}

As social VR has attracted various age groups, questions about social dynamics are raised. Maloney et al. (2020) \cite{Maloney2020a} used a participatory observation approach to gather 80+ hours of data from mainstream social VR platforms (AltspaceVR, RecRoom, VRChat). This allowed investigating the perspectives and interactions among minors and between minors and adults, observing nuanced interactions, positive and negative. From minor-minor interactions, themes captured were: 1) virtual intimacy and emotional connections via virtually-embodied non-verbal communication, trust and teamwork; 2) social interactions beyond gameplay, where older teenagers seemed to enjoy interacting more than playing with each other compared to younger teenagers; 3) sharing and disclosing in groups; and 4) dealing with harassment and bullying. Looking at adult-to-minor interactions, researchers noted mixed sentiments about each other, with the main themes being: 1) barriers, tensions and frustrations where adults would be 'annoyed' or 'frustrated' by minors' behaviours seeking attention; 2) mutual learning (relationship building, cultural learning); 3) social distancing from minors; and 4) adults discussing content inappropriate for children \cite{Maloney2020a}. While this study improved the understanding of social VR interactions between adults and minors, we extend their work significantly by collecting data directly from users who had a first-hand experience of interaction in social VR with minors, asking users about their feelings, actions taken afterwards and how it may have led to instigating new rules on their children or their own use. We also provide parents' perspectives and new insight into potential interventions for target age groups.

Social VR is a great opportunity to engage and interact with others in an innovative and exciting way, but it also creates problematic situations and encounters, especially with the presence of minors unsupervised, and leads to new forms of abuse, harm and harassment. A study that included 30 semi-structured interviews with adults (21 cis male, 5 cis female and 4 trans women) \cite{Maloney2020b}, focused on their social experiences in social VR, without directly asking them about interactions with children or adolescents. The main themes in the data included experiences, tensions and frustrations when encountering minors. For example, the prevalence of immature behaviours such as children screaming, negative social environments where young people are exposed to inappropriate behaviours and children's excessive use of social VR. Adults also expressed safety concerns for younger users, including the exposure of children to negative social environments and the excessive use of social VR, and worries about health and academic performance. In contrast to that work, our questionnaire focused exclusively on experiences in social VR that involve children.

Several recent events and studies have shown the emergence of harassment in social VR \cite{10.1145/3359202, Freeman2022}. Researchers conducted 25 semi-structured interviews (23 men and 2 women all from the US) and showed that virtual environments that allow embodiment and presence intensify online harassment experience \cite{10.1145/3359202}. This study resulted in the classification of harassment into: verbal (insults, hate speech via messages or voice), physical (unwanted touching via avatar movement), and environmental (sexual or violent content displayed via shared screens or virtual objects). As these social VR environments have little to no regulation, there are unclear norms of appropriate behaviours. 
We, therefore, focus on problematic events and negative situations in social VR to better understand the concerns and needs from a larger and diverse sample of adults' and parents' perspectives to create a safer social VR environment for children.

\subsection{Parenting and Technology}
\label{sec:parentingtech}

Children grow up in a digital world with ubiquitous connectivity, where the mobile phone and digital media (e.g., social media and online video games such as Massively Multiplayer Online Role-Playing Games (MMORPG) create means of connections and social interactions without necessarily their parents' knowledge and engagement. This has led to parental concerns. Prior work has investigated parents' perspectives, concerns, roles and influence on children's digital media and mobile phone use \cite{Odgers, Steinkuehler2016, ClarkBookChapter1, Radesky2016, Danet2020, Hartikainen2016, Boyd2007, Ito2018}. 

Parents tend to be concerned that digital media amplifies risk, including cyberbullying, children's social isolation, involvement in illegal activities, a threat to academic and physical development and mental health decline \cite{ClarkBookChapter1, Ito2018, Boyd2013}. A study questioned French parents (N = 149) of school-aged children, regarding their concerns and child's use of digital devices \cite{Danet2020}. Parents were mainly concerned that digital devices could harm the child's socio-emotional and cognitive development and they feared the technology would be at the expense of other activities. They were also concerned about the risk of being exposed to inappropriate content and the risk of addiction \cite{Danet2020}. One of the main difficulties parents have is controlling the children's content and screen time \cite{Danet2020, Odgers, Steinkuehler2016}.

Parental mediation, defined as "the parental management of the relationship between children and media`` \cite{Livinstone2008} has been considered a "new" type of parenting. Parents can shape and influence their children's media use and habits \cite{Radesky2016}. However, they can struggle with the unfamiliarity of technology, wanting greater transparency about their children's use of technology \cite{Blackwell2016} or their own use and attitudes towards the technology can impact and influence children's use \cite{Radesky2016, Smahelova2017}. Two approaches to parenting regarding digital media and devices were described: the ethic of "expressive empowerment" where parents emphasise the individual rights and trust of their child and the ethic of "respectful connectedness" where parents prioritise the rights of the family and familial goals \cite{ClarkBookChapter9}. Different rules, parental controls used and safeguarding practices vary depending on the children's ages, cultural background and parent-child relationships \cite{lorrie2014, Wisniewski2014}; and, as studies on social media showed, parent-child trust is crucial, and trust and control are not mutually exclusive \cite{lorrie2014, Hartikainen2016}. Existing means for child safety on the Internet and social media include: industry mediation (e.g., age limits), policies and educational efforts (e.g., schools), social mediation (e.g., parental active mediation) and technical mediation (e.g., parental controls) \cite{Hartikainen2016}. Moreover, parents may employ direct boundaries (limits and controls) or indirect boundaries (monitoring) with their child \cite{Erickson2016}. 

\subsection{Gap and Contribution over Prior Work}

In contrast to digital media, social VR increases the illusion of ``being there'' \cite{10.1145/3197391.3205451}. Social VR influences children’s inhibitory control and social behaviour compared to 2D screens \cite{bailey2019} and VR specifically influences children’s psycho-social development at different ages and stages due to its unique components compared to other 2D-screen devices \cite{Kaimara2021}. Moreover, controls and mitigation that exist in digital media (e.g., Microsoft Family Safety, Apple Families, Google Family Group) now have (largely) yet to be transposed to social VR \cite{Serkan2009, Costello2017, Ali2021, Clarke2019}. This motivated us to investigate in our questionnaire what mediation tools or strategies parents perceived to be needed for different children's age groups in social VR. While social VR can strengthen family relations when parents experienced social VR openly with their children \cite{Maloney2020a}, in other cases, teenagers mentioned that the technology would disconnect them from their family members. Moreover, prior research describes reasons why teenagers use social VR \cite{Maloney2021} but does not quantify the number of minors and the frequency of use. While participatory observations \cite{Maloney2020a} improved our understanding of social VR interactions between adults and minors, our work extends this by directly asking parents and non-parent adults about their feelings about interactions with children in Social VR, actions taken afterwards and the consequences on their children or their own use, as well as asking them to share notable first-hand experiences of encounters with children in social VR, interventions they currently use, and interventions they aspire to have.

\section{Online Mixed-Methods Questionnaire} 

The mixed-methods questionnaire was designed to gather information from parents and non-parent adults. We aimed to get a better understanding of their concerns, attitudes, encounters with children and willingness to use and adopt safety strategies. 
This will allow us to later develop parental mediation tools (i.e., approaches that provide parents with insight, oversight and/or control over the children's social VR experience) for these platforms. We used an online mixed-methods questionnaire (see supplementary material). The mixed-methods questionnaire was released in May 2022 on multiple forums (Reddit social VR and VR communities: VRGaming, VirtualReality, OculusQuest, VRchat, Oculus, RecRoom, AltspaceVR, metaverse; Oculus Community; AVForums; Twitter). This allowed us to target communities from a population of interest that has knowledge of VR and social VR to some extent. 

\subsection{Design and Method}
Employing the critical incident technique \cite{Flanagan1954}, we collected anonymous stories of social VR users of age 18 or more. 
Participants were asked to self-report a situation in social VR in which they encountered or interacted with minors without disclosing personal information.
Similar to recent studies on social media \cite{Huber2018} or educational VR \cite{Mado2022}, our approach entailed collecting data—currently lacking in the literature—from parents and legal guardians about their children’s use of social VR, safeguarding practices and concerns. 
Participants could optionally sign up for a lottery of one of two online shop vouchers. The study was approved by our ethics committee.

\subsection{Questionnaire Structure}

\begin{figure}[ht]
  \centering
    \includegraphics[width=1\columnwidth]{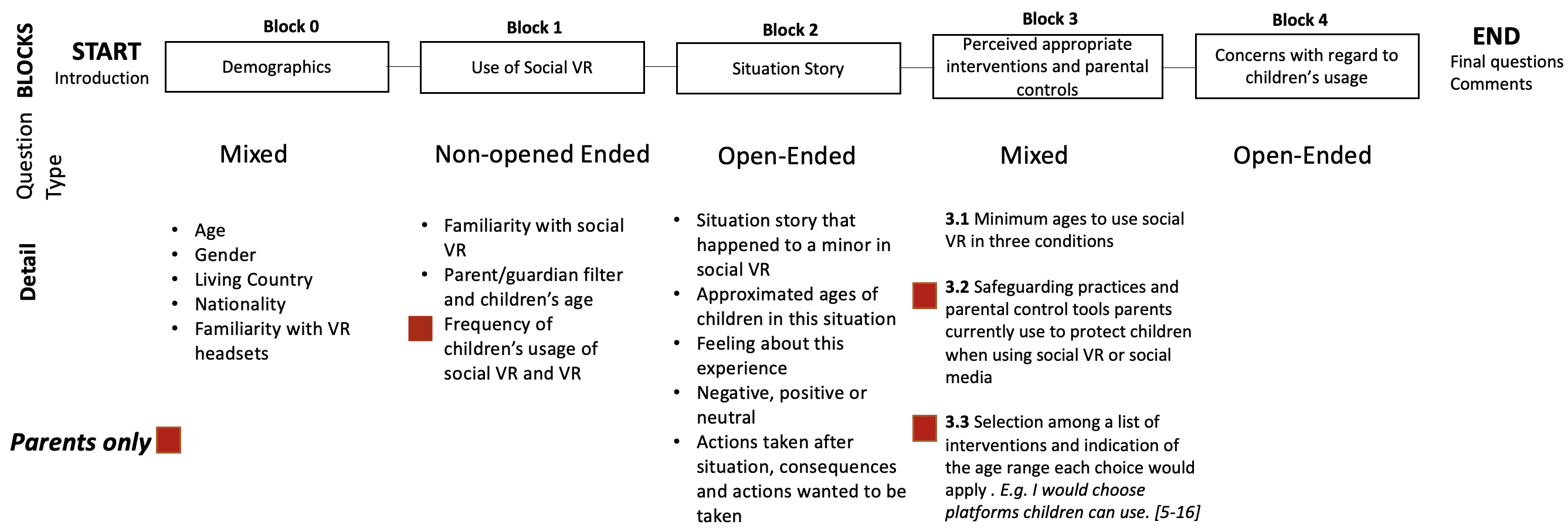}
    \caption{Mixed-methods questionnaire structure with information on question types and example items for each of the topics.}
    \label{figure:structure}
\end{figure}

The questionnaire was divided into five blocks with a mix of open-ended questions and non-opened-ended questions (sliders and Likert scales). Some questions were shown only to participants who indicated that they are parents or legal guardians of children. We will hereafter refer to this group as ``parents''. 
The structure of the questionnaire is illustrated in \autoref{figure:structure}.

Participants first completed standard demographic questions and were asked to indicate their familiarity with VR and social VR using a 6-point Likert scale (Block 0). 
In Block 1, participants were asked if they are guardians or parents of children that use VR or social VR, and some questions about the nature of children's use of social VR. 
In Block 2, we asked participants to write about a notable story that happened, without specifying in the question our focus on the negative aspects. Only after they answered the question, they were able to express how they felt and categorise the events or perceptions into negative, neutral or positive.
Block 3 was about the participants' perceptions of appropriate minimum ages for social VR (3.1), and of safeguarding practices they currently use (3.2), and would seek to use (3.3). The 18 interventions in Block 3.3 were chosen under four categories: pre-planned controls (e.g., education, setting up rules); knowledge of what happens as it happens (e.g., casting, receiving notifications); real-time interventions (e.g., blocking, muting) and after-the-fact (e.g., checking history). These were based on literature \cite{Freeman2022}, existing interventions used for social media \cite{Sciacca2022}, or potential interventions that could create a safer environment based on studies describing social VR issues and concerns \cite{10.1145/3359202}. The list of interventions is depicted in the Appendix (see supplementary material).

\subsection{Participants}
In total, 166 participants took part in the questionnaire, out of which 112 completed all of it. The questions addressed singular research outcomes (i.e., analysis of later questions was not predicated on analysis of prior questions). Therefore, we did not exclude partially completed responses as many of them were relevant and valid. The following responses however were considered invalid: repeated responses in multiple unfinished questionnaires and irrelevant responses (i.e., words or sentences not answering questions). We omitted participants when all response fields were empty, either starting from the demographics section or from block 2. This left us with 149 responses (Mean age (in years) = 32.59, SD age (in years) = 6.85).
Among participants, 38 are female, 107 are males, 2 are non-binary/third gender and 2 preferred not to say. 27 mothers, 51 fathers and 1 parent that did not disclose their gender responded to the questionnaire (i.e.,79 parents). Among the 70 non-parent participants there were 17 non-parent siblings or teachers.
Two participants are from Africa, 8 from Asia, 25 from Europe, 3 from Oceania, and 110 from the Americas (94 from the USA, 10 from Canada and 7 from South America). 
The participants’ age, knowledge and familiarity of general VR, social VR (6-point Likert Scale;  0=not at all; 5=extremely familiar) are summarised in \autoref{tab:demographics}.
Users reported the social VR platforms where the situation story occurred: 40 used VR Chat, 42 used Rec Room, 34 used AltspaceVR, 1 used Gorilla Tag, and 1 used Zepeto. Among parents, 15 parents were familiar with both VR and social VR and 23 were less familiar with either. Among non-parent adults, 15 were less familiar and 20 were familiar. 4 parents have never used social VR. 

\begin{table}[h]
\centering
\footnotesize
\begin{tabular}{llll}
  \toprule
\textbf{Groups} & & \textbf{Parents} & \textbf{\begin{tabular}[c]{@{}l@{}}Non-parent Adults\end{tabular}} \\
\midrule
  \textbf{N} & & 79 & 70 \\ 
  \midrule
  \textbf{Age (in years)} & MEAN [SD] & 32.59 [6.85] & 26.64 [8.20] \\ 
  \textbf{VR headsets Familiarity} & MEAN [SD] & 4.05 [0.80] & 4.06 [0.88]\\
   \begin{tabular}[c]{@{}l@{}} 0-5 Likert scale ('not at all' to 'extremely') \end{tabular} \\ 
  \textbf{Social VR familiarity} & MEAN [SD] & 3.88 [0.83] & 4.19 [0.90] \\
  \begin{tabular}[c]{@{}l@{}} 0-5 Likert scale ('not at all' to 'extremely') \end{tabular} \\
  \textbf{Children Age (in years)} & MEAN [SD] & 9.69 [4.39] & NA \\
   \bottomrule
\end{tabular}
\caption{Questionnaire Demographics Table Summary. (N = 149, 79 Parents, 70 Non-Parents). The questionnaire was mainly released on forums about VR/social VR, there may be self-selection bias, in which more engaged parents and non-parent adults in VR/social VR may have been more likely to participate in this study. However, this enabled us to get responses from users familiar with  VR/social VR.}
\vspace{-10mm}
\label{tab:demographics}
\end{table} 

\subsection{Analysis}
We analysed our quantitative data using R statistical tools (Blocks 0, 1, 3.1, 3.3) and our qualitative data with NVivo (Blocks 2, 3.2, 4). The data and the code used for the analysis can be found in the supplementary materials and will be made publicly available upon request.
The quantitative analysis involved data wrangling, summarising and visualising, and computing statistical tests.

Participant responses with regard to the situation story, concerns and safeguarding practices were coded using inductive thematic analysis \cite{Nowell2017}. This process consisted of the following four phases: data familiarisation, systematic data coding, generating initial themes by one coder and developing and reviewing themes and refining themes by three co-authors. For block 4 (Concerns about children’s usage), we reported results based on the questions asked and identified common concerns.
The generation of initial recurring themes was done in three steps. The first step involved a thorough reading of the collected data by the first author. The second step involved creating a spreadsheet and a second read of the qualitative data, taking notes throughout and summarising quotes into a few keywords. The third step involved the identification of initial topics and coding these with NVivo. All authors then helped refine the themes through multiple discussions over eight weekly meetings. 

\subsection{Limitations}

Our sample was not diverse in a number of aspects. Despite broad online recruitment on international websites, participants were mainly from the United States and two-thirds were Male, only 38 were from Africa, Europe, Asia and Oceania. This is in line with current VR usage trends \cite{statista}.  
The questionnaire was released in one language only (English) on forums that may not have allowed a more global reach. However, our sample reflects the demographics of VR.
Our participants are more engaged and familiar with social VR than the average user. While this allowed us to gather insights stemming from actual lived experiences, it also means that we do not capture novice experiences e.g., an experience of a user who tried social VR once and decided not to use it again. This is important to address in future work particularly to capture whether certain risks are amplified in these situations.
Approximately 50\% of our sample is a parent, allowing us to gather valuable insight for the purpose of this study. 
We also elected to seek stories from participants of age 18 or more for ethical reasons. 
While we focus on negative experiences and perceptions, we did not directly ask participants to tell us about negative experiences but rather a ``notable'' situation agnostic of valence. Accordingly, we did not explicitly ask about the benefits of social VR but rather left it open to respondents as to whether they would choose to report positive or negative experiences. Participants were asked to describe situations where they encountered or interacted with minors in social VR and to estimate their perceived ages. Consequently, the findings we report may on occasion be reflective of witnessed adult behaviours presented with a degree of immaturity. Subjective perceptions of age are nonetheless important and the best estimate available as respondents would rely on a variety of contextual cues (context, actions) for the age approximation. 
Moreover, the coding of the qualitative data was conducted by one researcher, and then the resultant codes and themes were discussed and refined by a team of three researchers.
Future work would require reaching out to a more global and diverse audience as VR continues to see increasing adoption outside of Western demographics and collecting children's perspectives on their social VR use and the interventions. 

\section{Results}

We first present findings on the children's use of social VR, looking at the frequency of use of social VR and VR in general by children's age groups (based on estimates reported by parents), as well as findings regarding what minimum ages are thought to be appropriate in social VR (4.1). 
We then present the qualitative findings with an overview of block 2 responses (4.2), focusing on our participants' experiences (4.3) and perceptions (4.4) of social VR towards adults and children. We also present non-parent adults' versus parents' concerns about social VR children's use (4.5). 
We describe safeguarding practices and tools used by parents (4.6). 
Finally, we show what interventions are selected by parents and the age ranges of children they think interventions are adapted for (4.7). 

\subsection{Parent-Reported Children's Usage of Social VR and Appropriate Minimum Ages for Social VR from Parent and Non-Parent Adult Perspectives}

\subsubsection{Frequency of Children's Social VR Usage and VR in General} 
\begin{figure}[!ht]%
    \centering
    \subfloat[\centering Frequency of Children's Use of VR in General]{{\includegraphics[width=6.5cm]{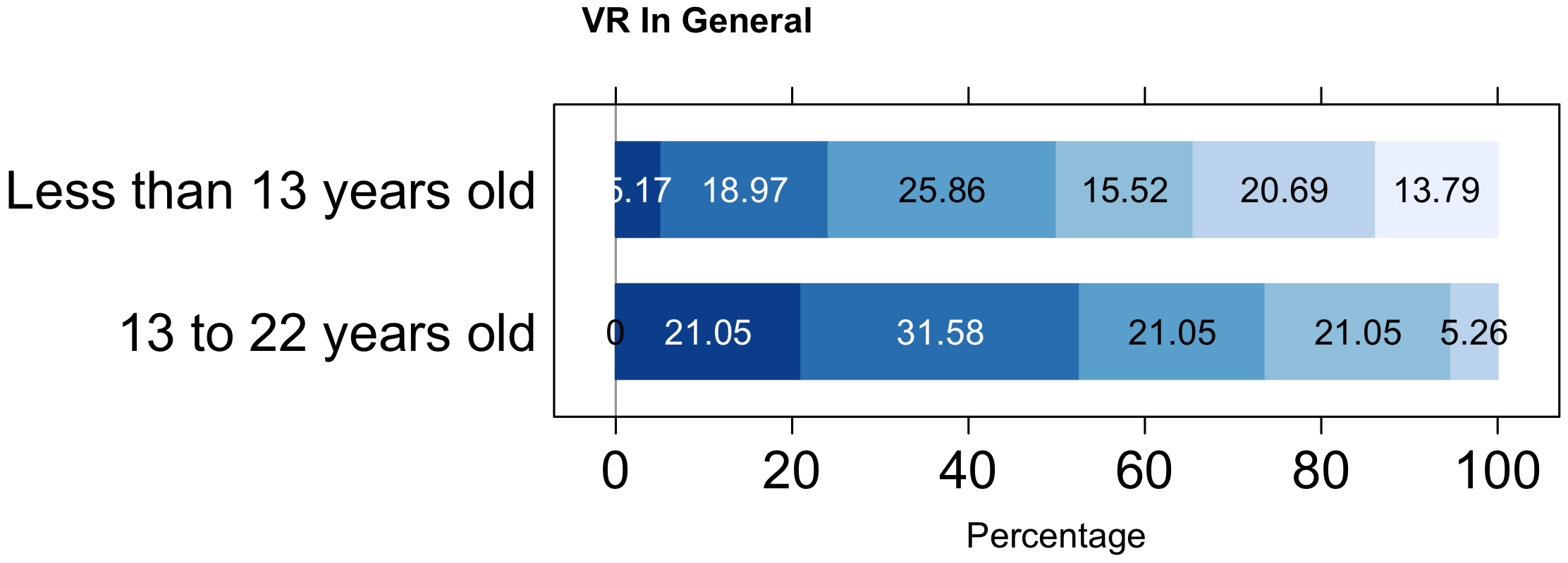} }}%
    \subfloat[\centering Frequency of Children's Use of Social VR]{{\includegraphics[width=6.5cm]{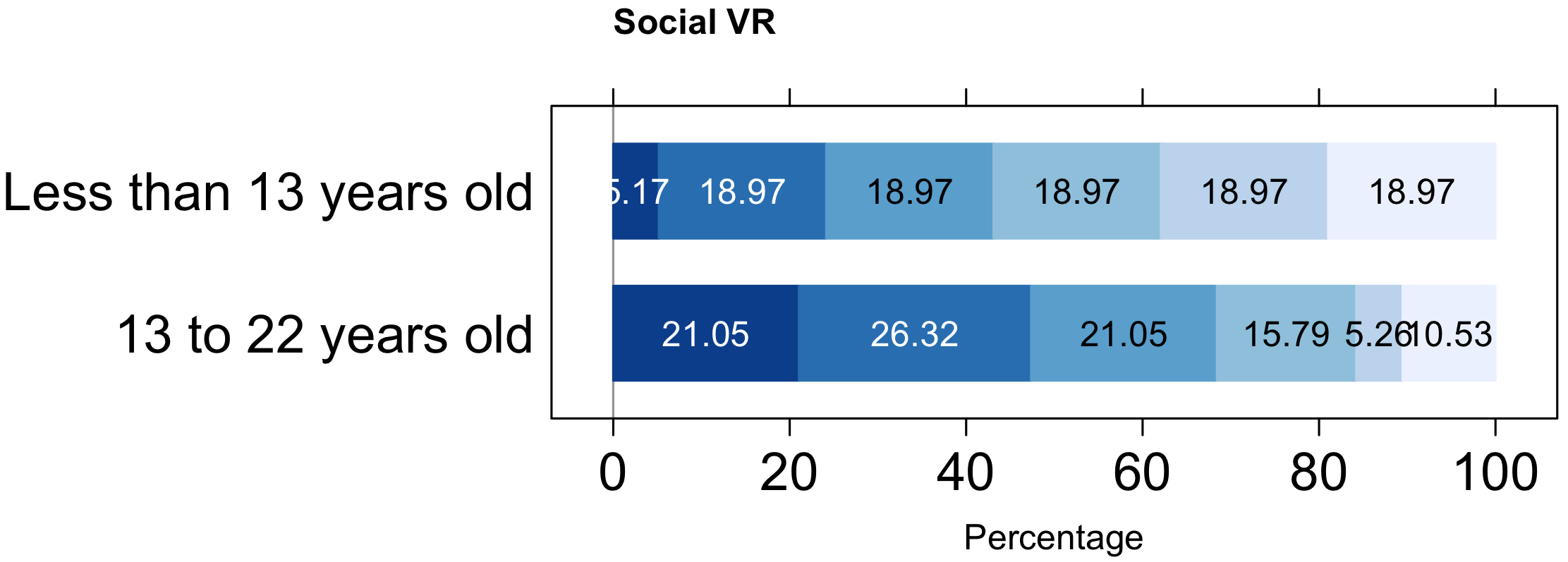} }}%
    \qquad
    \subfloat{\includegraphics[width=0.8\columnwidth]{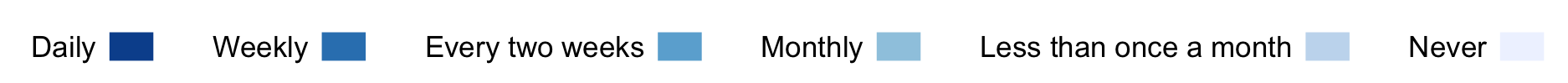}}%
    \qquad
    \caption{Likert Scales showing percentages of children using (a) VR in general or (b) Social VR in our sample. Among children under 13 years, \revisions{24\%} use social VR daily or weekly and \revisions{24\%} use VR in general daily or weekly. \revisions{43\%} use social VR at least every two weeks.}%
    \label{figure:likerts}%
\end{figure}

Parents reported their children's frequency of social VR use. We divided the sample into two age groups: parents of children under 13 years old \revisions{(N=58)} and between 13 (inclusive) and 22 years old \revisions{(N = 19)} and observed the frequency of use among participants’ children per age group. We chose these two groups based on current recommendations of VR and social VR platforms use, many of which require users to be 13 or older \cite{oculus, htc}. Among children of our participants that are less than 13 years old \revisions{(N=58)}, \revisions{43\%} use social VR at least every two weeks, whereas \revisions{50\%} use VR in general at least every two weeks. \autoref{figure:likerts} shows the frequency of VR and social VR use respectively in each group.

VR headset manufacturers have restricted the age limit to 13 years old+ and most manufacturers warn against its use for children \cite{oculus, htc}. However, we can clearly observe that many children under 13 use VR and social VR. Social VR platforms (e.g., VRChat, RecRoom) mostly recommend 13-year-olds or above and some provide junior accounts (e.g., RecRoom offers Junior Accounts for 9+ children) \cite{vrchat, recroom}. 
Industry mediation with age limits is not sufficient to restrain children's use as most of the platforms rely on self-professed age \cite{boyd_Hargittai_2010, Hartikainen2016}. Other types of mediation are therefore needed to safeguard and supervise children.

\finding{Despite age restrictions, some under 13-year-old children (\revisions{43\%} in our data) use social VR regularly (at least every two weeks) against the recommendation of manufacturers and developers.} \label{sec:finding1}

\subsubsection{Appropriate Minimum Age in Social VR from Non-Parent Adults' and Parents' Perspectives}

\begin{figure}[ht!]
    \centering
    \subfloat[\centering  Q1) What is the minimum age you think it is appropriate to use social VR where there can be adults and minors, without adult supervision?]{{\includegraphics[width=4cm]{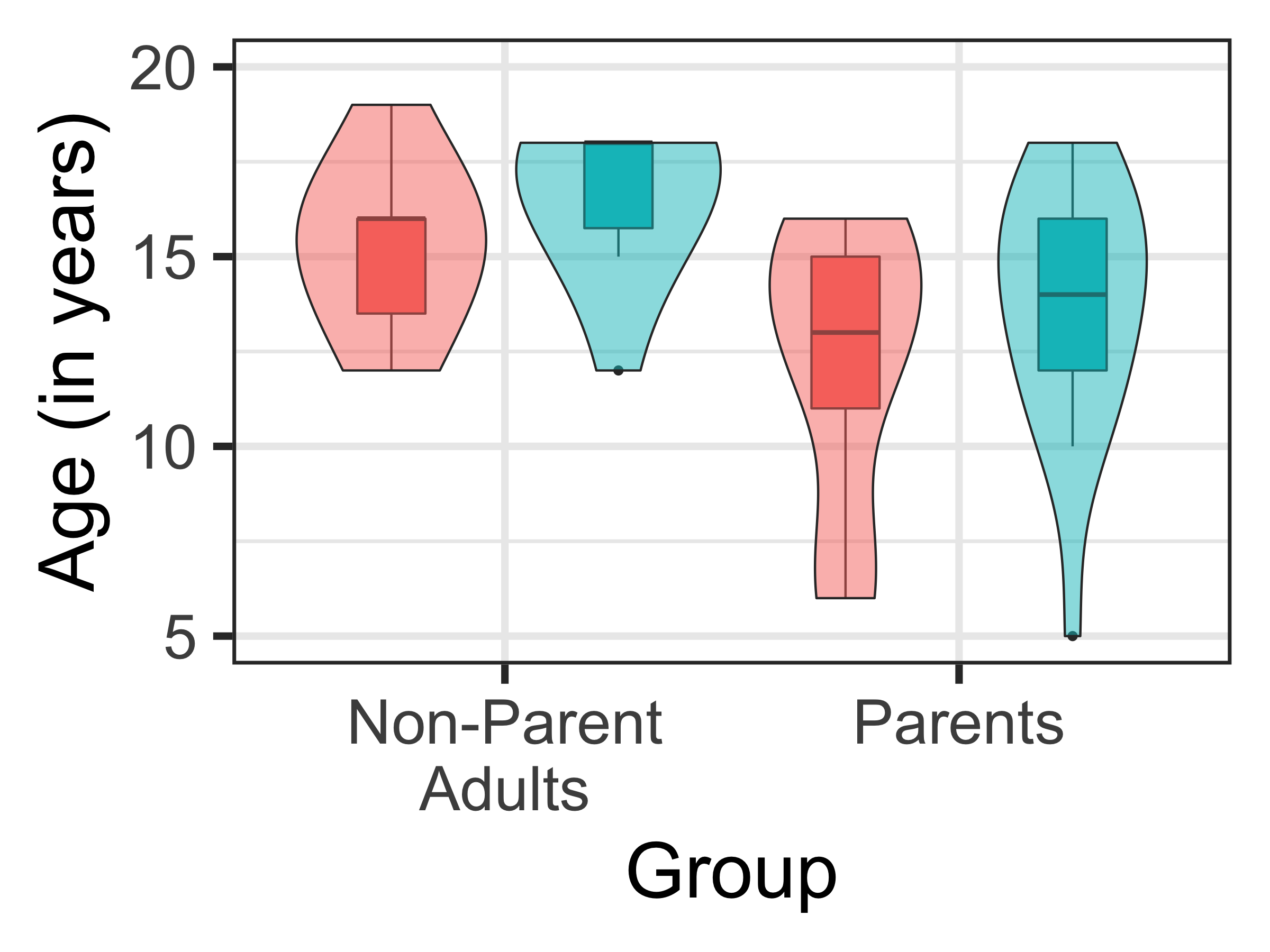} }}%
    \qquad
    \subfloat[\centering Q2) At what age do you think it is appropriate for a minor to use social VR given adult supervision?]{{\includegraphics[width=4cm]{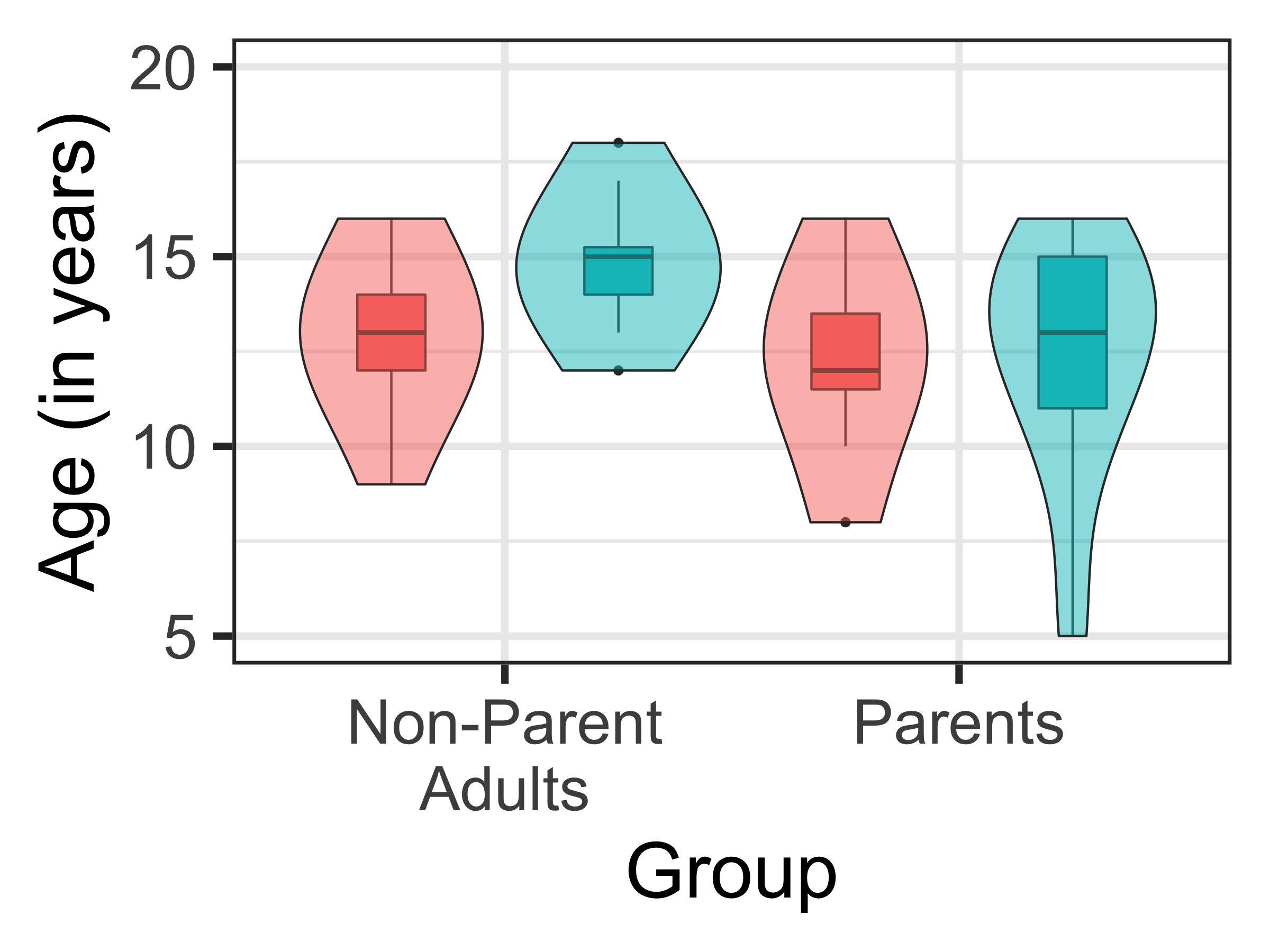} }}%
    \qquad
    \subfloat[\centering Q3) What is the minimum age you think it is appropriate to use social VR if there are only minors?]{{\includegraphics[width=4cm]{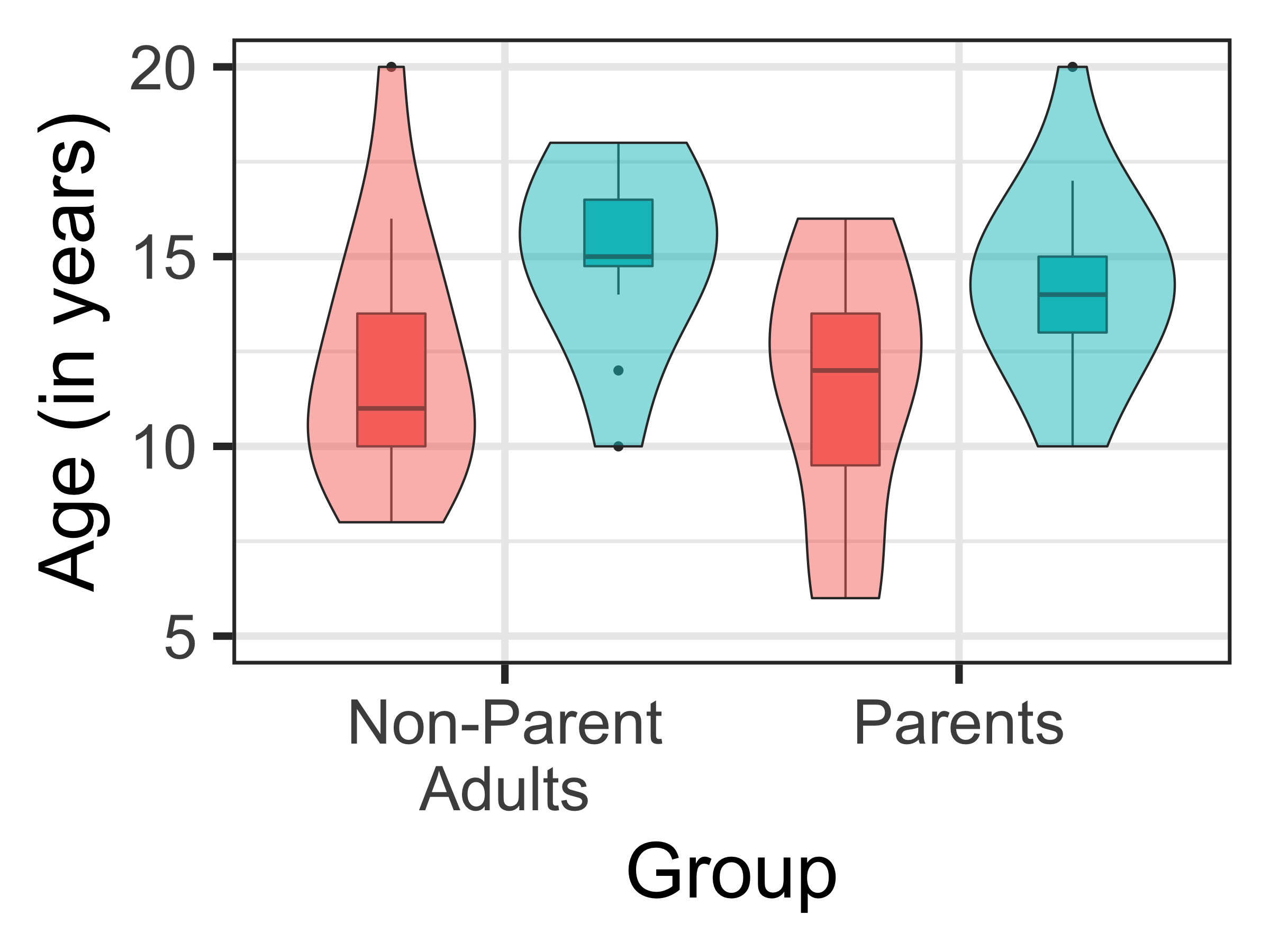} }}%
    \qquad
    \subfloat{\includegraphics[width=0.6\columnwidth]{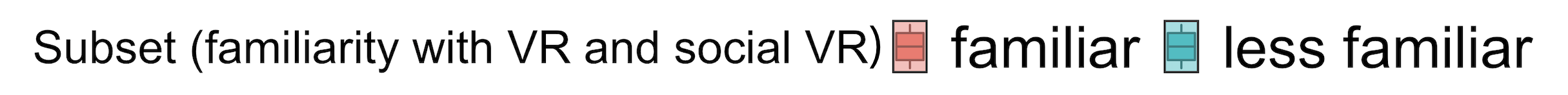}}%
    \vspace{-2mm}
    \caption{Violin and boxplots of inputted ages by parent and non-parent adult subsets corresponding to the minimum appropriate ages in social VR for three different supervision states (a, b, c). We conducted a three-way Aligned Rank Transform ANOVA (significant level 0.05) with age as the dependent variable and with parenthood groups (non-parent adult and parent), subsets (familiarity with VR and social VR) and questions (supervision states) as the factors involved. The main effect of familiarity with VR and social VR is statistically significant ($F$(1,54)=5.62, $p$= 0.022$<$0.05, $\eta_{p}^{2}$=.1, medium effect size). The main effect of parenthood is statistically significant ($F$(1,54)=5.60, $p$= 0.021$<$0.05, $\eta_{p}^{2}$=.1, medium effect size). The main effect of supervision states (Q1) without supervision, Q2) with supervision, Q3) only minors) is statistically significant ($F$(2,108)=13.49, $p$= 5.9e-06$<$0.01, $\eta_{p}^{2}$=.2, large effect size). The interaction between familiarity and supervision states is statistically significant
    ($F$(2,108)=6.14, $p$= 0.003$<$0.01, $\eta_{p}^{2}$=.1, medium effect size). The interaction between parenthood groups and supervision states is statistically significant ($F$(2,108)=5.03, $p$= 0.008$<$0.01, $\eta_{p}^{2}$=.1, medium effect size)} %
    \label{figure:age_appropriate}%
    \vspace{-3mm}
\end{figure}

While there are age recommendations from VR manufacturers and social VR platforms, children increasingly use social VR regardless of the guidelines \cite{Maloney2021}.
We asked participants what the appropriate age is for using social VR when 1) adults can interact with minors with supervision, 2) adults can interact with minors without supervision, and 3) minors can only interact with minors. We grouped results based on parenthood and familiarity with VR/social VR (\autoref{figure:age_appropriate}). 

After verifying assumptions of normality, homogeneity and extreme outliers to conduct a three-way ANOVA, the data did not pass the normality test (Shapiro-Wilk test). We, therefore, conducted a three-way Aligned Rank Transform (ART) ANOVA \cite{Elkin2021} with age as the dependent variable and with subset, group and question as the factors involved. Significant results (significant p-values set at 0.05) with medium (between 0.06 and 0.14) and large effect sizes (0.14 or higher) partial eta squared $\eta_{p}^{2}$ \cite{RICHARDSON2011135, Morris2013}, were obtained for parenthood groups (non-parent and parent) ($F$(1,54)=5.60, $p$= 0.021$<$0.05, $\eta_{p}^{2}$=.1, medium effect size), subsets (familiarity with VR and social VR) ($F$(1,54)=5.62, $p$= 0.022$<$0.05, $\eta_{p}^{2}$=.1, medium effect size) and questions (supervision states) ($F$(2,108)=13.49, $p$= 5.9e-06$<$0.01, $\eta_{p}^{2}$=.2, large effect size) with interaction effects also noted. In particular, from post-hoc comparisons, (1) parents familiar with VR and social VR are significantly more likely to let younger children use social VR without supervision compared to less familiar non-parent adults ($p$= 0.015$<$0.05). (2) Compared to parents and non-parent adults that are less familiar with VR and social VR, those familiar with them are significantly more likely to let younger children use social VR ($p$= 0.022$<$0.05). (3) Parents and non-parent adults are significantly more likely to let younger children use social VR with supervision ($p$$<$0.0001) and if there are only minors ($p$= 0.001$<$0.01) compared to those without supervision. 

\label{sec:finding2}
\finding{Perceptions of age appropriateness are influenced by parenthood, familiarity with social VR and supervision in social VR. Parents in our sample, in particular those familiar with VR/social VR, have the tendency to choose lower minimum ages for social VR usage compared to non-parent adults less familiar with either. Participants are more likely to let younger children use environments with supervision or where there are only minors, compared to without supervision.}

\subsection{Experiences with, and Perceptions Towards, Social VR Children and Other Adults}

Participants (N=149) were asked to recall and report instances from the following four types of encounters in social VR: 1) encountering users they suspect are children (60\% of N), 2) encountering users that were definitely children (62\%), 3) interacting (e.g., speaking) with children on social VR platforms (54\%), and 4) encountering children in social VR in areas that were not age-appropriate (40\%). Participants could select multiple options.
We asked participants to tell us about a notable situation that happened to a minor or minors they knew or encountered in social VR. If the situation was selected as a “Negative situation” (among "Negative", "Neutral" and "Positive'), they were asked what actions they took after the situation and what actions they would have taken or would have wanted to be taken. 
We collected 63 responses: \textbf{40} negative, \textbf{15} neutral, and \textbf{8} positive. Stories are reported in Section \hyperref[sec:stories]{4.3}.

Participants were also asked to approximate the ages of the children they met or encountered in social VR during the situation they described. Most of the participants estimated the age through the child's voice or height. The mean estimated age was younger than 13 (Mean = 11.42, SD = 3.33). 
After coding the described situations, the following two categories emerged: \textbf{experiences with children}, where participants described a one-off situation, and \textbf{perceptions/views and concerns} where participants expressed a sentiment of use or concerns about the presence of children in social VR. For both, participants described how adults or children were affected. 
Themes and counts of experiences and perceptions of said encounters are summarised in \autoref{table:matrix}. This model may allow us to better design interventions for different subjects and consider different perspectives, yet to be tested. 

\begin{table}[ht!]
  \centering
    \includegraphics[width=0.8\columnwidth]{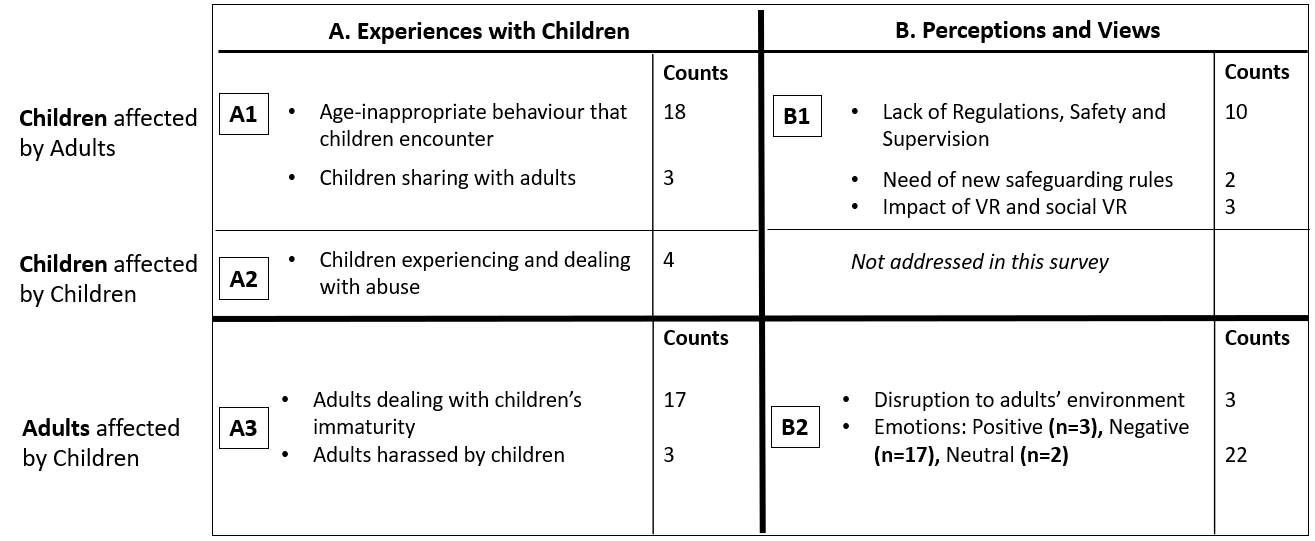}
    \caption{Four-Quadrant Matrix presenting themes from social VR adults and children experiences and perceptions from non-parent adults' and parents' perspectives. Themes and counts were obtained from coding the data which originated from free text answers. While the question asked about a one-off situation (A. experiences), participants also gave their perceptions and sentiments of use (B. perceptions and views). We do not consider "affected by adults" as it is out of our scope but may be considered in the future.
    \hyperref[sec:A1]{[Category A1.]} Experiences affecting children by adults. \hyperref[sec:A2]{[Category A2.]} Experiences affecting children by children. \hyperref[sec:A3]{[Category A3.]} Experiences affecting adults by children. \hyperref[sec:B1]{[Category B1.]} Perceptions and views towards children
    \hyperref[sec:B2]{[Category B2.]} Perceptions and views towards adults.}
    \label{table:matrix}
    \vspace{-5mm}
\end{table}

\subsection{Notable Experiences Encountering Children in Social VR}
Themes around experiences with children raised by both parents and non-parent adults are depicted in this section. They are generated from answers involving children as the main subjects, affecting children or adults. The count of responses on experiences categorised into themes is given in \autoref{table:matrix}.

\subsubsection{Experiences that Affect Children in Social VR (A1, A2)}
\label{sec:stories}
Responses in this category were grouped into three themes: children encountering age-inappropriate behaviours, children facing abuse, and children self-disclosing to adults.

\paragraph{Age Inappropriate Behaviours That Children Encounter}
\label{sec:A1}

Participants described situations in social VR where children were exposed to inappropriate content and conversations (e.g., sexual graphic content, erotic play, conversations that include drugs, sex and alcohol), forms of oppression (e.g., violence, racism, homophobia and verbal abuse) or had easy access to age-inappropriate environments/areas (e.g., VR nightclubs). 
The following is an example of inappropriate situations children can be exposed to and also points to the high likelihood of encountering children in social VR:

\begin{quote}
\textit{``It's an extremely common occurrence to see children in these environments exposed to the sexually risqué avatars and actions of others, either directly or indirectly. In many cases, children will either start or join into these experiences on their own volition. [...] With the added obfuscation a headset provides in seeing these age inappropriate situations, it is very easy for a child to get into them without parental knowledge. As such, you would be hard pressed to *not* find a child in a lobby. Based on personal experience, I have never been in a Social VR lobby without a child present somewhere.``} [non-parent adult] (P13).
\end{quote}

Participants also mentioned children were being treated as adults or children acted as adults in social VR, for example:
\begin{quote}
\textit{``People talking trash at a pool table [...] and one of the players revealing they were only 14. The mood changed and some were a bit nicer but the random minor playing has their feelings hurt and was treated like an adult (harshly).``}[parent] (P31)
\end{quote}

\begin{quote}
\textit{``Someone I knew several months ago (I have them blocked for now until they turn 18) was a minor that was lying about their age and engaging in very much adult things (drinking and sexually charged activity) for several months before when they confessed to it. They said they felt trapped and peer pressured into acting older after people initially assumed that they were of age.``} [non-parent adult] (P35)
\end{quote}

\paragraph{Children Experiencing and Dealing with Abuse}
\label{sec:A2}

Children would experience forms of abuse in social VR, similar to those in the real world, such as bullying by other children or verbal abuse from teenagers and young adults. Some children may ignore them but others may be affected emotionally and physically or may want to act the same way e.g.:

\begin{quote}
\textit{``My son frequently interacts with other people in social games who sound like late teens or young adults who use graphic sexual and homophobic language in "trolling". They're annoying and he's good at ignoring them. It reduces his enjoyment of the games, but it's hard to see how to stop this. [...] He usually plays where I can hear him and we talk about any abusive players and how to handle it.``} [parent] (P6)
\end{quote}

\paragraph{Children Sharing with Adults} In social VR, children may encounter adults that would be friendly and supportive of what they are going through or wish to share. This may be beneficial for children who may not want to share with someone they personally know. While some adults might support children by only listening to them, children may not be cautious and disclose personal information to malicious adults. This result is in line with current findings showing the prevalence of self-disclosure in social VR \cite{Sykownik2022}.

\begin{quote}
\textit{``I have also had conversations with a few kids about how they are happy that I talk to them [...]. They feel heard when people just listen to what they have to say. There was one kid that was going through a time when his parents were divorcing and this was a way for him to get away from that. He talked about crying some nights wishing it never happened. [...]``} [non-parent adult] (P12)
\end{quote}

\subsubsection{Experiences that Affect Adults in social VR (A3)}
\label{sec:A3}

Experiences affecting adults were grouped into two themes: adults dealing with children's immaturity and adults being harassed by children.

\paragraph{Adults Dealing with Children's Immaturity} Adults face children's immaturity in social VR environments initially designed for adults and young adults. Adults typically complained about children's behaviour in social VR. Children create disruptive environments in \textit{"mature crowd in public lobbies"} [parent] (P1). They like to be inappropriate and children use these environments as a way to push boundaries:
\begin{quote}
    \textit{``It feels like an outlet for children to misbehave or test the boundaries without facing real consequences.``} [non-parent adult] (P61)
\end{quote} Children shout, yell and run around (e.g., a child started crying when a [non-parent adult] participant (P18) won a game).
Participants mentioned other adults encouraging children to misbehave and adults shaming children. However, adults typically attempted to mitigate these disruptions. They would deal with these disruptions via: social distancing (adults leaving the VR rooms when in presence of children or actively avoiding children and not engaging with them), blocking children and adults having to turn down the volume and muting others. For example:

\begin{quote}
\textit{``One yelled/screeched at me for spawning next to them. It was a little unpleasant. [...] I turned my volume all the way down because I’m not using headphones and my housemates can hear them. Once the sound was off, I navigated through the menu to leave the room.``} [non-parent adult] (P56)
\end{quote}

Among answers dealing with disruptions, adults also mentioned paying attention to their behaviour by watching their language in front of children and trying to be nice.

\paragraph{Adults Harassed by Children}
Adults may also be verbally harassed by children, for example: 
\begin{quote}
    \textit{``One example of many is seeing a sexually risqué avatar controlled by an adult being verbally harassed by children.``} [non-parent adult] (P13)
\end{quote}

\finding{1) Both parent and non-parent adult social VR users in our sample have to deal with abuse and immaturity by children towards them. 2) Children are often exposed to age-inappropriate behaviours and activities, they face abuse, and sometimes use Social VR to seek moral support and self-disclose to adults.}
\label{sec:finding3}

\subsection{Parent and Non-Parent Adult Perceptions about Youth in Social VR}

\subsubsection{Non-Parent Adult and Parent Perceptions Towards Children in Social VR (B1)}
\label{sec:B1}
Participants gave their views about children being present in social VR and raised their concerns and worries towards children (Count = 18). In particular, participants reported a lack of regulation and supervision in social VR, and how children may not know how to protect themselves and navigate safely in social VR. Responses also described the effects and impacts of VR and social VR on children, these were subsequently grouped into negative and positive. Negative impacts included physical (e.g., eyestrain) and mental health impact (e.g., mental development, trigger of negative emotions, negative influence). Examples of negative impacts that are unique to social VR included: adopting or learning poor behaviour and becoming out of touch with social reality.
Positive effects included the possibility of children having fun and being fully immersed and teenagers willing to learn, socialise and introduce social VR to their friends. 

\begin{quote}
\textit{``Concerned that a younger user might stumble into something inappropriate for them without realising it or meaning to. [...] Reminded me that many children play the game and makes me hope that more of them have good supervision/are willing to leave when people encourage them in case things get weird.``} [non-parent adult] (P34)
\end{quote}

\begin{quote}
\textit{``Minors [under] 13 years of age should [be] under adult supervision and must ensure intermittent use and rest. Long-term use should be avoided, as possible negative effects include: imbalance of hand-eye coordination, affecting the brain's ability to multi-task, and adults should closely monitor the use of this product by minors.``} [parent] (P64)
\end{quote}

However, parents felt social VR could be comparable to real-world situations where parents have to trust their children in situations that may not always have supervision, for example:

\begin{quote}
\textit{``I think this constitutes standard exploratory behaviour by teenagers of adult environments, and with the added barrier of pseudonymity and non-physical interaction from VR it is a less dangerous or simply real situation than e.g. someone in their mid-teens going to a house party.``} [parent] (P65)
\end{quote}

\subsubsection{Non-Parent Adult and Parent Perceptions Towards Adults in Social VR (B2)}
\label{sec:B2}
While some responses were about how social VR can affect children as described above, other responses noted how the presence of children in these environments affected adults (Count = 6) and how children would need supervision (Count = 12). Adults face discomfort, annoyance, verbal abuse and harassment from children:

\begin{quote}
\textit{``I find these experiences deeply disturbing as I don't believe these children should have access to a VR headset without strict supervision. The sheer amount of swearing, insults and racial slurs is astonishing both directed at and from these children. As my own children grow up I will not be allowing them access to a VR headset without strict supervision. These games allow unfiltered contact between children and adults which leads to inappropriate communication.``} [parent] (P3)
\end{quote}

\begin{quote}
\textit{``You see and hear children on platforms, such as VRChat where the minimum age is 13. Why do parents give a young child access to an expensive piece of technology and put them on a platform with an age requirement that they fail to meet?''} [non-parent adult] (P51)
\end{quote}

Based on Russell's Circumplex model polarities \cite{Russell1980}, we categorised emotions felt by users. For instance, negative emotions would include anger, frustration and stress and positive emotions would include happiness, serenity and relaxation. Perceptions were mostly negative (17 negative emotions, 2 neutral emotions and 3 positive emotions). 
Our analysis indicates that the feelings of users who noticed being observed were negative in the majority of cases (e.g., 'worry, concerned' (P36, P43, P75, P101, P139), 'bothered' (P3), 'disturbing' (P5), 'inappropriate' (P5, P36), 'annoyed' (P75), 'angry' (P87), 'embarrassed' (P58), 'isolated' (P99), 'dependent' (P99), 'felt bad' (P33, P98), 'sad' (P34, P103), 'surprised, shocked' (P118, P138). 
Some participants felt both 'happy and sad' (P14, P18), 'normal' (P26) or 'neutral' (P67). 
P104 felt 'happy' and P107 'good'. 

\begin{quote}
\textit{``I felt bad for the kid but rejection is part of hanging out with new people. I think lobbies should be sorted out by relative player age so kids don’t have to experience things they aren’t ready for.``} [non-parent adult] (P31)
\end{quote}

\finding{Non-parent adults and parents in our sample typically perceive the presence of children in social VR as negative, affecting both children and adults negatively. They cite the lack of safeguarding and regulation in social VR spaces.}
\label{sec:finding4}

\subsection{Parent and Non-Parent Adult Concerns about Children's Social VR Use}

Both parents and non-parent adults are concerned about the freedom children have in these virtual environments where there can be malicious people (e.g., paedophiles) (Count = 8) and the lack of supervision, regulations and monitoring that can lead to several consequences: psychological (Count = 28) (e.g., addiction, mental health issues, skewed self-image, decreased real-world social awareness and learning skills) and physical (Count = 12) (e.g., eye-strain, impact on the body, anxiety). 
Parents expressed their concerns (Count = 33) with a particular focus on nurturing their parent-child relationship and wanting to build trust, for example:

\begin{quote}
\textit{``I'm no more concerned about VR than the internet in general. Maybe even less so, as almost all use in my house happens in front of me (due to space constraints, not a rule). Cultivating trust with your child and speaking openly about the rewards and dangers of online interaction is more important than surveillance (which can backfire and erode trust).``} [parent] (P6)
\end{quote}

Among non-parent adults' concerns (Count = 34), a common concern referred to parents' lack of awareness regarding all the possibilities of interactions (both positive and negative) and the extent to which children are taking advantage of this freedom in social VR:
\begin{quote}
\textit{``[...] I think social VR can be a wonderful tool and a lot of fun, even for younger users, but I think that there are not really a lot of safeguards available for them at the moment on most platforms.``} [non-parent adult] (P34)
\end{quote}

\begin{quote}
\textit{``I am worried about the difference in knowledge between children and their parents about what occurs on VR platforms. It seems like there is a lot more freedom, and a lot more possibilities of interaction (both positive and negative) than parents believe.``} [non-parent adult] (P166)
\end{quote}
\label{sec:finding5}
\finding{1) Parents in our sample prioritise trust and conversations whereas non-parent adults in our sample prefer enforcing rules. 2) Major concerns regarding children's use of social VR revolve around the issue of excessive freedom and psychological negative impacts (e.g., addiction, worries about children losing real-world social skills and learning abilities).}
\label{sec:finding5}

\subsection{Safeguarding Online: Practices and Tools}
\label{sec:safeguarding}

\subsubsection{Actions Taken, Consequences or Actions Wanted to Be Taken}

We investigated what actions participants took after the event they encountered and what actions they would have liked to be taken in these similar events. 
Actions taken mostly involved active communication, explanation and guidance. Both non-parent adult and parent social VR users explained to children involved in the situations what went wrong soon after the incidents. Parents mentioned limiting or reducing their child's social VR usage and communicated about the impact of these inappropriate behaviours and disruptive situations before incidents. 
We also asked their views on what should be done and what they would have wanted to be done. One of the most wanted actions from non-parent adults is age verification, age restriction and limitations such as allowing minors to login only in the weekends. They also want improved moderation, education and suggested that developers should create safe places and protection and supervision systems. They expect parents to better moderate child social VR use.
Parents would like strong authentications and time limits. A parent (P39) mentioned children should not be allowed somewhere unattended, at least until they know how they can process the information and know how to stop. Parents also expect more guidance and supervision of children in social VR.

\begin{quote}
\textit{``Minors should be accompanied by guardians.``} [non-parent adult] (P132) \end{quote}

\begin{quote}
\textit{``Developers, publishers, and platforms should find ways to enforce their terms of service and make sure that safe places are safe, and open places are open. I do not blame the kids.``} [non-parent adult] (P61)
\end{quote}

\begin{quote}
\textit{``Parents are always monitoring their children. Before the children have no ability to distinguish themselves, parents still play the role of enlightenment teachers. In view of current VR, it is suggested that parents choose VR equipment and content reasonably. [...]``} [parent] (P163)
\end{quote}

Regarding consequences, parents have restricted and limited children's social VR use, stop children from using VR completely, watch and monitor:

\begin{quote}
\textit{``It used to be five times a week, but now it's once a week."} [parent] (P45)
\end{quote}

\finding{1) Parents in our sample want robust authentication and time limits. They stress that children should not be unsupervised. More guidance and supervision for children on these platforms is expected. 2) Non-parent adults in our sample strongly want age verification, age restrictions and improved moderation for social VR platforms. They urge developers to create safe spaces with better protection and supervision. They also expect parents to be more proactive in moderating their children's usage.}
\label{sec:finding6}

\subsubsection{Parents' Safeguarding Practices}
\label{sec:safeguarding2}

We asked parents how they monitor or safeguard their children who use social VR and what parental or privacy controls they use for social media for example. 
Parents' answers on safeguarding included two main types: real-world safeguarding practices (external to the device, Count = 28) and virtual-world safeguarding practices (internal, via the device, Count = 12). 
The first type (real world) involved communication and education (i.e., education about anti-violence, communication on safety and privacy, rules, parents and guardians participating with children) and physical space and time limits (i.e., children permitted to use VR in the living room only, physical space safety to avoid children to hurt themselves when they use head-worn devices, presence of parents when they use VR, limiting the time of use).

\begin{quote}
\textit{``He usually uses it in the living room where I can hear what's going on. We have built trust around his online activities, he knows he can come to me to talk about anything that might bother him.``} [parent] (P5)
\end{quote}

The second type of safeguarding (VR world) involved: checking the history, watching over (e.g., via streaming), authentication, and downloading suitable applications. Tools and existing controls they use included: blocking games/sites/rooms, privacy controls, surveillance, time limit checks, monitoring accounts history, muting and blocking, and mobile phone monitoring, social media privacy controls, mobile phone to stream VR content.
\label{sec:finding7}
\finding{Parents in our sample use a variety of safeguarding practices, both in the real-world (Count = 28) to put physical and time boundaries in place, but also through the virtual world (Count = 12) with existing tools that mostly include the mobile phone to stream VR content or controls for checks and authentication.}

\subsection{Parents' Interventions Selection and Corresponding Age Range}
\label{sec:interventions}

\begin{figure}[h]
    {{\includegraphics[width=0.7\columnwidth]{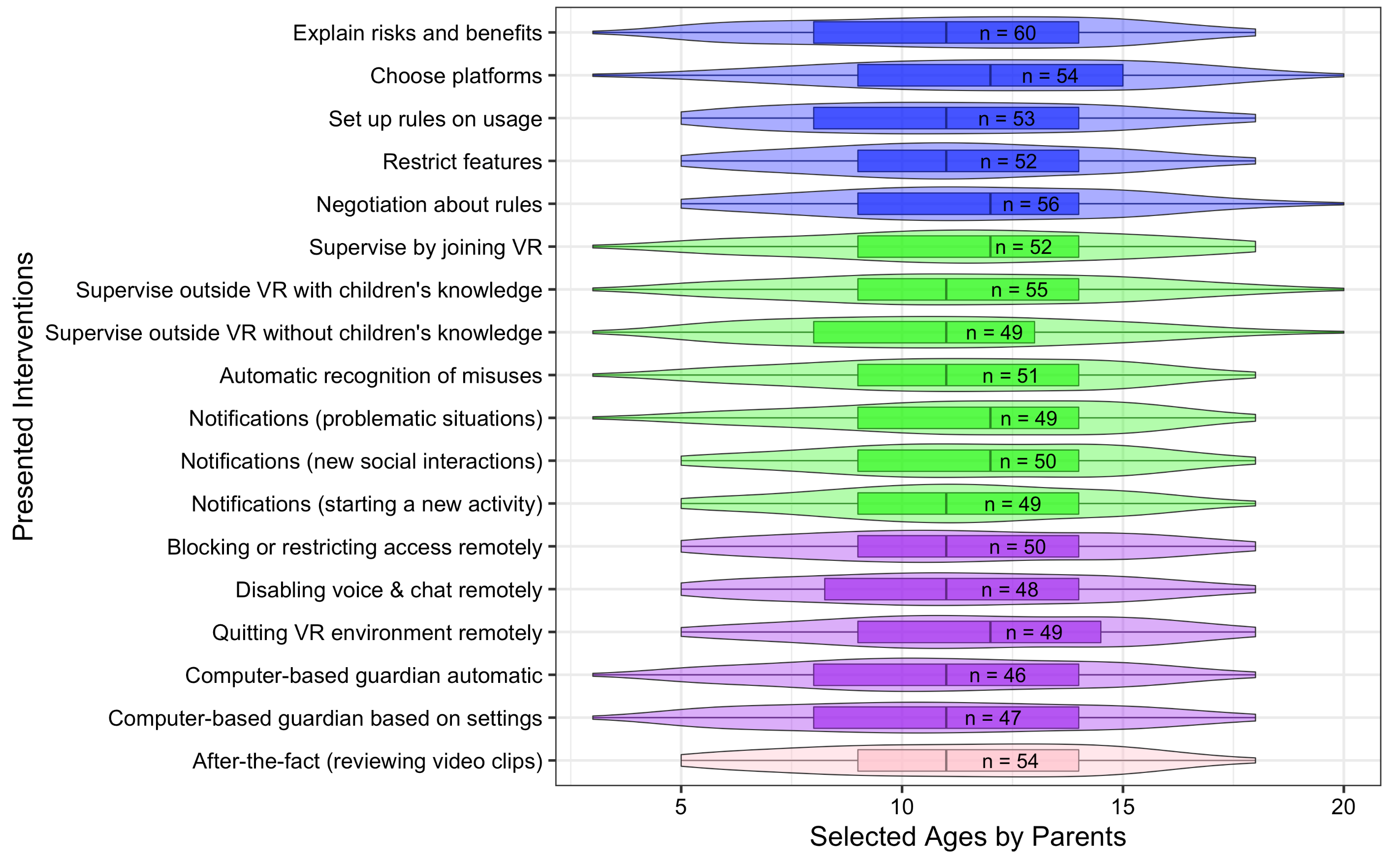}}}
    {{\includegraphics[width=0.13\columnwidth]{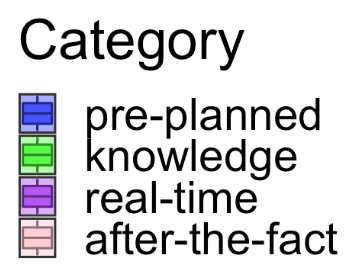}}}
    \caption{Interventions presented to the participants and the selected age range. The violin boxplots summarise age ranges selected for four categories of interventions (pre-planned, knowledge, real-time actions and after-the-fact). Each intervention was selected by n participants. The mean of minimum ages given for all interventions is 10 years old and the mean of maximum ages is 15 years old.}
    \label{figure:boxplots}
\end{figure}

In this section, we summarise responses to questions in which parents were asked to 1) select from among 18 interventions those they would put in place (list of 18 interventions in Supplementary Material), 2) give an age range (from minimum to maximum age) for which they think the selected intervention is adapted for. We argue this list is an interesting slice of the intervention space but may need to be refined in future work. The interventions were grouped into four categories: pre-planned (e.g., choosing the content and platforms), knowledge (e.g., monitoring via notifications received), real-time actions (e.g., blocking remotely) and after-the-fact (e.g., reviewing the history (\autoref{figure:boxplots}).

\textbf{1) Proportions of selected interventions} We did not find significant differences between the number of selected interventions ($p$ $>$ 0.05 Chi-Square Test). 

\textbf{2) Ages given for interventions} Age ranges given by parents among all intervention categories went from 0 years old up to 20 years in the categories of knowledge and pre-planned interventions. While there were no significant results that showed a difference between the number of times an intervention was selected, interventions in all four categories were generally selected with a mean age range of 10-15 years old. We detect outliers (maximum ages given of 20 years old) for interventions in the category of 'knowledge as it happens' and 'pre-planned'. While these extremities are outliers, we argue that there could be a potential misuse of interventions.

Looking at minimum ages only, we found a significant difference between the minimum ages given for interventions ($F$(17,752)=1.68, $p$= 0.041$<$0.05, $\eta_{p}^{2}$=.04, small effect size, ART ANOVA). In particular, the minimum ages given for the intervention "Explain risks and benefits" were significantly lower than for the intervention "Negotiation about rules" ($p$= 0.02$<$0.05, Contrasts with Tukey adjustment). Moreover, the children's ages of parents in our sample did not have an effect on results as we may expect different opinions from parents who currently have younger children compared to those with older children. With interventions grouped into four categories, we observed significant differences between the minimum ages of pre-planned interventions and knowledge ($p$= 0.01$<$0.05, Contrasts with Tukey adjustment), with no effect from the parents' children's ages. 

\finding{Parents in our sample seek to apply a breadth of safeguarding practices and interventions to protect their children, with limited evidence of adaptation based on minimum ages given but no evidence of adaptation based on age ranges. This suggests the need for guidance around what interventions are suitable for developmental age ranges.} \label{sec:finding8}

\section{Discussion and Future Work}

\subsection{Main Findings}
We conducted an online mixed-methods questionnaire about children's usage of social VR collecting data from the ecosystem adult stakeholders of child social VR users. A summary table of the findings, methods and participants can be found in the Supplementary Material. The findings align with previous research \cite{Maloney2020a, Maloney2020b, Maloney2021, Sykownik2022} but also contribute new perspectives, highlighting both advancements and knowledge gaps. Our questionnaire both confirms the transferability of some results from research on media literacy and parenting \cite{Odgers, Steinkuehler2016, ClarkBookChapter1, Radesky2016, Danet2020, Hartikainen2016, Boyd2007, Ito2018}, and presents novel insights that are specific to Social VR. Indeed, concerns towards social VR also include psychological and physical risks (e.g., addiction, being exposed to inappropriate content and social isolation). Both parents and non-parent adults raised concerns regarding the increased immersion in harassment and bullying situations, the freedom children have in these virtual environments and the current lack of mitigation and supervision tools. However, as Boyd and Hargittai (2013) found \cite{Boyd2013}, while most parents express a high level of concern regarding children's online safety risks, levels of concerns can vary with background and demographic factors. 
As for safeguarding practices, our results exemplify a combination of the two approaches to parenting regarding digital media: "expressive empowerment" and "respectful connectedness" as described in \hyperref[sec:parentingtech]{[Section 2.3]}.
A full table summarising the findings can be found in the Supplementary Material.


\textit{\textbf{RQ1) Social VR Frequency and Appropriate Age of Children}} In the data, children less than 13 years old \revisions{(N=58)} use social VR and VR regularly with approximately \revisions{43\%} using social VR daily, weekly or every two weeks \hyperref[sec:finding1]{[Finding 1]}. This result emphasises the importance of developing interventions to protect and safeguard children in social VR environments initially designed for adults. Our findings also showed the impact of familiarity with VR and social VR, parenthood and supervision conditions in which minors would use social VR (without supervision, with supervision and only with minors) on what participants would think is an appropriate age to use social VR \hyperref[sec:finding2]{[Finding 2]}.
\textit{\textbf{RQ2) Parent and Non-parent Experiences Influenced by Minor’s Use of Social VR}}
Among parents, we observed that those who are familiar with VR and social VR are significantly more likely to rate bonding and tension at home higher due to social VR usage than parents who are less familiar with social VR or VR.
Themes obtained from our coding were similar to those obtained in Maloney's and his colleagues' work \cite{Maloney2020a} including the following: "dealing with harassment and bullying" (in minor-minor interactions), "social distancing from minors" and "adults discussing inappropriate for children" (in adult-minor interactions). Our findings depict concrete experiences affecting adults or affecting children to perceptions and concerns towards adults or children \hyperref[sec:finding3]{[Finding 3]}. In the data, most non-parent adults were concerned about the lack of supervision and safeguarding tools and the freedom children have in these environments, with mentions of how parents fail to meet age limits recommendations. 
There was an overwhelming number of negative answers and negative emotions compared to those that were positive or neutral based on Russell's Circumplex model \cite{Russell1980} \hyperref[sec:finding4]{[Finding 4]}.
\textit{\textbf{RQ3) How Parents Choose to Moderate and Limit the Social VR Experience}}
Parents’ answers were polarised, some would practice strict supervision or completely stop their child from using social VR due to their negative experiences and perceptions, and the lack of regulation and intervention tools. Others would consider social VR similar to real situations where teenagers experiment and may not be under supervision when going to a party, requiring them to build a parent-child trusting relationship \hyperref[sec:finding5]{[Finding 5]}, \hyperref[sec:finding6]{[Finding 6]}, \hyperref[sec:finding7]{[Finding 7]}. 
Parents in the data seek to apply a range of interventions with little adaption based on age. The minimum ages given for the intervention "Explain risks and benefits" were significantly lower than for the intervention "Negotiation about rules". The vast majority of interventions were applied indiscriminately with respect to age ranges, not taking into account different developmental stages \hyperref[sec:finding8]{[Finding 8]}. Proactive parents engaged to protect their children from risks are seen as "good" parents \cite{Boyd2013}. However, as Glassner (1999) reports \cite{Glassner1999}, people misassess risk, especially in new environments where they have little experience, thus, leaning towards increased restrictions. It is therefore important to understand parental concerns when developing social VR interventions and regulations but also consider children's perspectives, maturity and age.

\subsection{Addressing Negative Interactions in Social VR: Guiding the Adult-Minor Dynamic}

Based on our findings and reflections, informed by social VR-aware parents and non-parent adults, we propose future directions for tackling negative interactions between adults and minors in social VR to create safer virtual environments.

\textbf{\textit{Age-Matching and Private Social Virtual Spaces}} Non-parent adults recommended higher minimum ages compared to parents \hyperref[sec:finding2]{[Finding 2]}. They raised concerns about the lack of parental supervision, the need for enforcement of rules and age restrictions in virtual environments where children enjoy considerable freedom \hyperref[sec:finding5]{[Finding 5]}, \hyperref[sec:finding6]{[Finding 6.2]}. To tackle these challenges, social virtual platforms could consider introducing private spaces that allow users to interact only with their friends with their own rules. However, it is important to be aware that such spaces may also introduce the risk of new forms of abuse, such as adults pretending to be, or forming connections with, child users to join these private spaces. Additionally, implementing age verification measures to create age-matching virtual spaces that users can join in addition to public spaces is an option also proposed in recent work \cite{Deldari2023}. This would necessitate users sharing their identification or using ID authentication technology \cite{Pasquale2020}, however, potentially giving rise to privacy, potential leakage and ethical issues \cite{Phippen2022, blogpost}. Future research is needed to better understand how age verification and inclusion of users from various age groups in group instances can be effectively implemented in the context of social VR while tackling the limitations.

\textit{\textbf{The Need for Behavioural Control and Access Control (Pre-Planned)}} Industry mediation with age limits is currently not sufficient to restrain children’s use as most of the platforms rely on self-professed age. While we may want to introduce biometrics to exclude children completely based on their voices or heights, for example, we argue that children may still be able to use social VR, the data may be fraught with errors and they may find ways around the mechanisms introduced oftentimes with parental consent and support \cite{Wilson2018}. Other types of mediation are needed to safeguard and supervise children in social VR to make experiences safer for children. Social VR environments currently lack mature technical mediation, and there is an almost complete absence of supervision and safeguarding tools for parents on most platforms. However, parents in our sample have been using 'Behaviour Control' (i.e., based on defined rules and procedures) \cite{Hartikainen2016}, in this case, by limiting children’s use and 'Clan Control' (i.e., acceptable behaviour is reinforced) \cite{Hartikainen2016}, by communicating and explaining the impact of these disruptive situations, or by restricting their child from using social VR \hyperref[sec:safeguarding]{[See Section 4.6.2]}. These controls are either done in the real world or via the virtual world. Regarding what participants would have wanted to moderate these disruptive situations, responses mainly included technical mediation for Access Control, in particular, age verification, login at the weekends only, secure authentication, protection and supervision systems \hyperref[sec:safeguarding]{[see Section 4.6.1]}.

\textit{\textbf{The Need for Oversight (Knowledge and Real-Time Interventions)}}
Parents also want interventions to oversee (i.e., real-time monitoring) their children in social VR to understand what is happening in the virtual spaces. Parents may not be able to monitor by co-using social VR with their child via another VR headset. As an alternative, some parents use casting or stream the virtual environment via their mobile phones but this requires them to focus on the video \hyperref[sec:safeguarding2]{[See Section 4.6.2]}. If they do not pay attention, they may miss notable events. Parents need ways to become aware of problematic situations, when their child is meeting someone or when they are starting a new activity. These alerts would facilitate gaining knowledge about what is happening in the environment. Parents also expressed their desire to be able to take an action remotely without having to use a headset. However, we found that in our sample, parents less familiar with VR and social VR are more cautious than those familiar with them, in particular, if there is no supervision \hyperref[sec:finding2]{[Finding 2]}. Therefore, oversight might be preferred by parents who are not aware of what usually happens in social VR and would want to get real-time notifications and interventions remotely such as blocking or muting. 

Recent work has proposed the implementation of automated (embodied) moderation taking into account the unique affordances of VR has and facilitate parental supervision (informing parents of negative encounters instantaneously), as an alternative and more child-friendly safety tool \cite{fiani2023IDC} to enforce rules, combat harassment and negative encounters in social VR \cite{schulenberg2023towards, fiani2023IDC}.

\textit{\textbf{The Need for Insight After-the-Fact}}
As previously mentioned, parents may not always be able to follow the video cast when their children use social VR. Therefore, the possibility of reviewing the child's experience asynchronously (e.g., through video recordings) after the event would facilitate parental insight into social VR. 

Parents have selected a wide range of interventions in our questionnaire with little adaption based on age \hyperref[sec:finding8]{[Finding 8}]. They mentioned wanting to build parental trust when moderating and safeguarding \hyperref[sec:safeguarding]{[See Section 4.6.2]}. Trust echoes prior research in other technologies \cite{Hartikainen2016}, highlighting the importance of parent-child trust. 

\textit{\textbf{Ethical Considerations Related to Parental Monitoring}} According to the Information Commissioner's Office (ICO) \cite{ICO}, children must have the right to know if their parents are monitoring them online, with an "obvious sign to the child" \cite{ICO}. Transparency is paramount. However, parents still selected the intervention "Supervise from outside VR without children's knowledge" with the maximum age mean being 15 years old.
Outliers were detected as age ranges were given with upper limits of 20 years old for interventions in the categories of 'knowledge' and 'pre-planned'. This result demonstrates potential misuse of interventions; for example, adults may use them without considering the best interest of their child, or without the child's consent. Therefore, we must consider exploitation risks and privacy concerns when developing interventions, ensuring these are in line with guidance from the ICO Children’s Code and the five rights (5Rights framework \cite{5rights}), regarding a child’s evolving right to privacy and freedom of expression. There is also the potential that such interventions could be misused by adults applying them to other adults (e.g., for domestic abuse) \cite{Tanczer2021, Slupska2021}, thus, care needs to be taken as to what interventions are provided. 

\textit{\textbf{The Need for Evidence-Based Guidance around Age-Appropriateness of Social VR Interventions}} Parents were unable or unwilling to apply interventions in a discriminant manner based on age \hyperref[sec:interventions]{[See Section 4.7]}. If social VR is to become a dominant means of child social interaction, it will also become a significant influence on social development, much as social media has impacted current generations \cite{Valkenburg2022}. Our finding affirms the need for evidence-based guidelines to support parents and platforms around age-appropriate interventions, to carefully balance safeguarding against developmental needs. We can argue there is a need for multi-disciplinary research involving Child-Computer Interaction experts and developmental psychologists.
Indeed, research shows the importance of developing different interventions for children of different age groups as each group has their unique development needs \cite{Freeman2022, Beals2009}. Parents' main role should be to ensure the protection of their children \cite{Budd2001}, but they may not be able to differentiate which interventions are best for a specific age group and which allow both control and trust. Future work is required to help them choose the most appropriate and effective mediation tools for specific developmental age ranges, maturity and developmental abilities. Parents also have different parenting strategies as the data showed, some parents would stop their child from using social VR completely whereas others would consider letting their child use social VR similar to letting their child go to a party. It is therefore important for developers to take into account age groups and facilitate parental insight and selection of interventions for appropriate age groups considering parent wants and child needs.

\textit{\textbf{Potential Obstacles: Tension at Home and Lack of Trust}}
The rise of technology such as smartphones and social media has shown to be a source of connection but also led to tension in families \cite{Maloney2020b}. While children may tend to overestimate the opportunities and benefits, they underestimate the risks \cite{Schofield2012}. However, some children feel like technology can create tension with the example of a boy not liking when his father texts on his phone while the boy plays soccer \cite{Schofield2012}. While parents are typically concerned about their children's social VR use and the potential risks, some may want to overly restrict social VR use which would impede parent-child trust. Future work is necessary to better understand parent-child relationships and how social VR may affect tension and bonding at home.

\textit{\textbf{Perspectives and Parent-Child Relationship}} As shown in previous work and our data, children’s safety in social VR is an issue that requires attention and parents' involvement. We must also consider children's perspectives on the interventions to allow a parent-child trusting relationship and allow parents to intervene without taking away the child’s sense of agency in the situation. We acknowledge that the choices of interventions can depend on parental strategies, cultures and children's ages. We plan to carry out such research with a more diverse and larger sample, with the exploration of participants' backgrounds including ethnicity, socio-economical status and education, to better understand what types of supervision are needed in social VR according to parents and children and to study interventions' satisfaction and effectiveness from both perspectives, via co-design methods.

\section{Conclusion}
Through our mixed-methods questionnaire, we found the extent to which minors use social VR, how parents’ and non-parent adults' experiences and perceptions of social VR are influenced by children's use and how parents would choose to moderate and limit children's social VR experience. No other studies have specifically focused on relating parental perspectives towards this disruptive technology, relevant for the development of parental and safety-enhancing tools. In our data, \revisions{43\%} of children under 13 use social VR at least every two weeks. We discovered that social VR familiarity, use of supervision and parenthood influence the perceptions of age appropriateness in social VR which open the discussion of cautiousness, parental responsibility and parent-child trust. Children face abuse, age-inappropriate behaviours and self-disclose to adults. Both parent and non-parent adult social VR users deal with children's immaturity and can be harassed by children. Participants substantially pointed out the lack of supervision, regulation and safety which leads to the disruption of adults' VR environment and has a negative impact on children. Parents seek to apply a breadth of safeguarding practices in the virtual or real world, restrictions, and interventions to protect their children, with little adaptation based on age. Therefore, parents need guidance on choosing appropriate and effective mediation tools for specific developmental age ranges, abilities and maturity. This work expands prior work, our understanding of negative adult-child interactions in social VR and outlines significant research challenges around the design of age-appropriate interventions and the need for the development of better technologies which may require the involvement of psychologists, parents, children and developers to enable effective parenting and protect children in social VR. 



\begin{acks}
We thank our participants. This work was supported by the UKRI Centre for Doctoral Training in Socially Intelligent Artificial Agents, Grant Number EP/S02266X/1, by REPHRAIN: The National Research Centre on Privacy, Harm Reduction and Adversarial Influence Online under UKRI grant: EP/V011189/1, and partly sponsored by a 2020 Meta Research Award on Responsible Innovation. 
\end{acks}

\bibliographystyle{ACM-Reference-Format}

\begin{thebibliography}{75}


\ifx \showCODEN    \undefined \def \showCODEN     #1{\unskip}     \fi
\ifx \showDOI      \undefined \def \showDOI       #1{#1}\fi
\ifx \showISBNx    \undefined \def \showISBNx     #1{\unskip}     \fi
\ifx \showISBNxiii \undefined \def \showISBNxiii  #1{\unskip}     \fi
\ifx \showISSN     \undefined \def \showISSN      #1{\unskip}     \fi
\ifx \showLCCN     \undefined \def \showLCCN      #1{\unskip}     \fi
\ifx \shownote     \undefined \def \shownote      #1{#1}          \fi
\ifx \showarticletitle \undefined \def \showarticletitle #1{#1}   \fi
\ifx \showURL      \undefined \def \showURL       {\relax}        \fi
\providecommand\bibfield[2]{#2}
\providecommand\bibinfo[2]{#2}
\providecommand\natexlab[1]{#1}
\providecommand\showeprint[2][]{arXiv:#2}

\bibitem[5ri(2018)]%
        {5rights}
 \bibinfo{year}{2018}\natexlab{}.
\newblock \bibinfo{title}{5Rights}.
\newblock
\newblock
\urldef\tempurl%
\url{https://5rightsframework.com/}
\showURL{%
\tempurl}
\newblock
\shownote{Last Accessed: 09-08-2022}.


\bibitem[sta(2022)]%
        {statista}
 \bibinfo{year}{2022}\natexlab{}.
\newblock \bibinfo{booktitle}{\emph{Advertising \& Media Markets -- AR \& VR}}.
\newblock
\urldef\tempurl%
\url{https://www.statista.com}
\showURL{%
Retrieved December 10, 2022 from \tempurl}


\bibitem[Rec(2022)]%
        {RecRoomSafety}
 \bibinfo{year}{2022}\natexlab{}.
\newblock \bibinfo{title}{Comfort and Safety — Rec Room}.
\newblock
\newblock
\urldef\tempurl%
\url{https://recroom.com/safety}
\showURL{%
\tempurl}
\newblock
\shownote{Last Accessed: 30-08-2022}.


\bibitem[vrc(2022)]%
        {vrchat}
 \bibinfo{year}{2022}\natexlab{}.
\newblock \bibinfo{title}{Community Guidelines — VRChat}.
\newblock
\newblock
\urldef\tempurl%
\url{https://hello.vrchat.com/community-guidelines}
\showURL{%
\tempurl}
\newblock
\shownote{Last Accessed: 22-07-2022}.


\bibitem[Alt(2022)]%
        {AltspaceVRsafety}
 \bibinfo{year}{2022}\natexlab{}.
\newblock \bibinfo{title}{User safety and moderation - AltspaceVR | Microsoft Docs}.
\newblock
\newblock
\urldef\tempurl%
\url{https://docs.microsoft.com/en-us/windows/mixed-reality/altspace-vr/user-safety}
\showURL{%
\tempurl}
\newblock
\shownote{Last Accessed: 30-08-2022}.


\bibitem[VRC(2022)]%
        {VRChatTrust}
 \bibinfo{year}{2022}\natexlab{}.
\newblock \bibinfo{title}{VRChat Safety and Trust System}.
\newblock
\newblock
\urldef\tempurl%
\url{https://docs.vrchat.com/docs/vrchat-safety-and-trust-system}
\showURL{%
\tempurl}
\newblock
\shownote{Last Accessed: 30-08-2022}.


\bibitem[htc(nd)]%
        {htc}
 \bibinfo{year}{n.d.}\natexlab{}.
\newblock \bibinfo{title}{Health and Safety | HTC United States}.
\newblock
\newblock
\urldef\tempurl%
\url{https://www.htc.com/us/sunrise-safety-guide/}
\showURL{%
\tempurl}
\newblock
\shownote{Last Accessed: 22-07-2022}.


\bibitem[mmo(nd)]%
        {mmostats}
 \bibinfo{year}{n.d.}\natexlab{}.
\newblock \bibinfo{title}{MMO Stats — Population \& Player Count of Online Games}.
\newblock
\newblock
\urldef\tempurl%
\url{https://mmostats.com/}
\showURL{%
\tempurl}
\newblock
\shownote{Last Accessed: 19-11-2022}.


\bibitem[ocu(nd)]%
        {oculus}
 \bibinfo{year}{n.d.}\natexlab{}.
\newblock \bibinfo{title}{Oculus Safety Centre | Oculus}.
\newblock
\newblock
\urldef\tempurl%
\url{https://www.oculus.com/safety-center/}
\showURL{%
\tempurl}
\newblock
\shownote{Last Accessed: 22-07-2022}.


\bibitem[rec(nd)]%
        {recroom}
 \bibinfo{year}{n.d.}\natexlab{}.
\newblock \bibinfo{title}{A Parent's Guide to Rec Room — Rec Room}.
\newblock
\newblock
\urldef\tempurl%
\url{https://recroom.com/parents-guide}
\showURL{%
\tempurl}
\newblock
\shownote{Last Accessed: 22-07-2022}.


\bibitem[ICO(nd)]%
        {ICO}
 \bibinfo{year}{n.d.}\natexlab{}.
\newblock \bibinfo{title}{Use of parental controls | ICO}.
\newblock
\newblock
\urldef\tempurl%
\url{https://ico.org.uk/for-organisations/childrens-code-hub/how-to-use-our-guidance-for-standard-one-best-interests-of-the-child/children-s-code-best-interests-framework/use-of-parental-controls/\#recommendations}
\showURL{%
\tempurl}
\newblock
\shownote{Last Accessed: 09-08-2022}.


\bibitem[sta(nd)]%
        {statsvrchat}
 \bibinfo{year}{n.d.}\natexlab{}.
\newblock \bibinfo{title}{VRChat (App) · Steam Charts · SteamDB}.
\newblock
\newblock
\urldef\tempurl%
\url{https://steamdb.info/app/438100/graphs/}
\showURL{%
\tempurl}
\newblock
\shownote{Last Accessed: 19-11-2022}.


\bibitem[Ali et~al\mbox{.}(2021)]%
        {Ali2021}
\bibfield{author}{\bibinfo{person}{Suzan Ali}, \bibinfo{person}{Mounir Elgharabawy}, \bibinfo{person}{Quentin Duchaussoy}, \bibinfo{person}{Mohammad Mannan}, {and} \bibinfo{person}{Amr Youssef}.} \bibinfo{year}{2021}\natexlab{}.
\newblock \showarticletitle{Parental Controls: Safer Internet Solutions or New Pitfalls?}
\newblock \bibinfo{journal}{\emph{IEEE Security and Privacy}} (\bibinfo{year}{2021}).
\newblock
\showISSN{15584046}
\urldef\tempurl%
\url{https://doi.org/10.1109/MSEC.2021.3076150}
\showDOI{\tempurl}


\bibitem[Bailey et~al\mbox{.}(2019)]%
        {bailey2019}
\bibfield{author}{\bibinfo{person}{Jakki~O. Bailey}, \bibinfo{person}{Jeremy~N. Bailenson}, \bibinfo{person}{Jelena Obradović}, {and} \bibinfo{person}{Naomi~R. Aguiar}.} \bibinfo{year}{2019}\natexlab{}.
\newblock \showarticletitle{Virtual reality's effect on children's inhibitory control, social compliance, and sharing}.
\newblock \bibinfo{journal}{\emph{Journal of Applied Developmental Psychology}}  \bibinfo{volume}{64} (\bibinfo{year}{2019}), \bibinfo{pages}{101052}.
\newblock
\showISSN{0193-3973}
\urldef\tempurl%
\url{https://doi.org/10.1016/j.appdev.2019.101052}
\showDOI{\tempurl}


\bibitem[Ball et~al\mbox{.}(2021)]%
        {BALL2021}
\bibfield{author}{\bibinfo{person}{Christopher Ball}, \bibinfo{person}{Kuo-Ting Huang}, {and} \bibinfo{person}{Jess Francis}.} \bibinfo{year}{2021}\natexlab{}.
\newblock \showarticletitle{Virtual reality adoption during the COVID-19 pandemic: A uses and gratifications perspective}.
\newblock \bibinfo{journal}{\emph{Telematics and Informatics}}  \bibinfo{volume}{65} (\bibinfo{year}{2021}), \bibinfo{pages}{101728}.
\newblock
\showISSN{0736-5853}
\urldef\tempurl%
\url{https://doi.org/10.1016/j.tele.2021.101728}
\showDOI{\tempurl}


\bibitem[Barreda-Ángeles and Hartmann(2022)]%
        {BARREDAANGELES2022}
\bibfield{author}{\bibinfo{person}{Miguel Barreda-Ángeles} {and} \bibinfo{person}{Tilo Hartmann}.} \bibinfo{year}{2022}\natexlab{}.
\newblock \showarticletitle{Psychological benefits of using social virtual reality platforms during the covid-19 pandemic: The role of social and spatial presence}.
\newblock \bibinfo{journal}{\emph{Computers in Human Behavior}}  \bibinfo{volume}{127} (\bibinfo{year}{2022}), \bibinfo{pages}{107047}.
\newblock
\showISSN{0747-5632}
\urldef\tempurl%
\url{https://doi.org/10.1016/j.chb.2021.107047}
\showDOI{\tempurl}


\bibitem[Beals and Bers(2009)]%
        {Beals2009}
\bibfield{author}{\bibinfo{person}{Laura Beals} {and} \bibinfo{person}{Marina~Umaschi Bers}.} \bibinfo{year}{2009}\natexlab{}.
\newblock \showarticletitle{A Developmental Lens for Designing Virtual Worlds for Children and Youth}.
\newblock \bibinfo{journal}{\emph{International Journal of Learning and Media}}  \bibinfo{volume}{1} (\bibinfo{date}{2} \bibinfo{year}{2009}), \bibinfo{pages}{51--65}.
\newblock
Issue 1.
\urldef\tempurl%
\url{https://doi.org/10.1162/IJLM.2009.0001}
\showDOI{\tempurl}


\bibitem[Blackwell et~al\mbox{.}(2019)]%
        {10.1145/3359202}
\bibfield{author}{\bibinfo{person}{Lindsay Blackwell}, \bibinfo{person}{Nicole Ellison}, \bibinfo{person}{Natasha Elliott-Deflo}, {and} \bibinfo{person}{Raz Schwartz}.} \bibinfo{year}{2019}\natexlab{}.
\newblock \showarticletitle{Harassment in Social Virtual Reality: Challenges for Platform Governance}.
\newblock \bibinfo{journal}{\emph{Proc. ACM Hum.-Comput. Interact.}} \bibinfo{volume}{3}, \bibinfo{number}{CSCW}, Article \bibinfo{articleno}{100} (\bibinfo{date}{nov} \bibinfo{year}{2019}), \bibinfo{numpages}{25}~pages.
\newblock
\urldef\tempurl%
\url{https://doi.org/10.1145/3359202}
\showDOI{\tempurl}


\bibitem[Blackwell et~al\mbox{.}(2016)]%
        {Blackwell2016}
\bibfield{author}{\bibinfo{person}{Lindsay Blackwell}, \bibinfo{person}{Emma Gardiner}, {and} \bibinfo{person}{Sarita Schoenebeck}.} \bibinfo{year}{2016}\natexlab{}.
\newblock \showarticletitle{Managing expectations: Technology tensions among parents and teens}.
\newblock \bibinfo{journal}{\emph{Proceedings of the ACM Conference on Computer Supported Cooperative Work, CSCW}}  \bibinfo{volume}{27} (\bibinfo{date}{2} \bibinfo{year}{2016}), \bibinfo{pages}{1390--1401}.
\newblock
\showISBNx{9781450335928}
\urldef\tempurl%
\url{https://doi.org/10.1145/2818048.2819928}
\showDOI{\tempurl}


\bibitem[Boyd and Hargittai(2010)]%
        {boyd_Hargittai_2010}
\bibfield{author}{\bibinfo{person}{Danah Boyd} {and} \bibinfo{person}{Eszter Hargittai}.} \bibinfo{year}{2010}\natexlab{}.
\newblock \showarticletitle{Facebook privacy settings: Who cares?}
\newblock \bibinfo{journal}{\emph{First Monday}} \bibinfo{volume}{15}, \bibinfo{number}{8} (\bibinfo{date}{Jul.} \bibinfo{year}{2010}).
\newblock
\urldef\tempurl%
\url{https://doi.org/10.5210/fm.v15i8.3086}
\showDOI{\tempurl}


\bibitem[Boyd and Hargittai(2013)]%
        {Boyd2013}
\bibfield{author}{\bibinfo{person}{Danah Boyd} {and} \bibinfo{person}{Eszter Hargittai}.} \bibinfo{year}{2013}\natexlab{}.
\newblock \showarticletitle{Connected and concerned: Variation in parents' online safety concerns}.
\newblock \bibinfo{journal}{\emph{Policy \& Internet}}  \bibinfo{volume}{5} (\bibinfo{date}{9} \bibinfo{year}{2013}), \bibinfo{pages}{245--269}.
\newblock
Issue 3.
\showISSN{1944-2866}
\urldef\tempurl%
\url{https://doi.org/10.1002/1944-2866.POI332}
\showDOI{\tempurl}


\bibitem[Boyd(2007)]%
        {Boyd2007}
\bibfield{author}{\bibinfo{person}{Danah~M. Boyd}.} \bibinfo{year}{2007}\natexlab{}.
\newblock \showarticletitle{Why youth (heart) Social network sites: the role of networked publics in teenage social life}.
\newblock \bibinfo{journal}{\emph{MacArthur Foundation Series on Digital Learning – Youth, Identity, and Digital Media}}  \bibinfo{volume}{7641} (\bibinfo{year}{2007}), \bibinfo{pages}{1--26}.
\newblock
Issue 41.
\showISBNx{9780262524834}
\showISSN{15292401}
\urldef\tempurl%
\url{https://doi.org/10.1162/DMAL.9780262524834.119}
\showDOI{\tempurl}


\bibitem[Budd(2001)]%
        {Budd2001}
\bibfield{author}{\bibinfo{person}{Karen~S Budd}.} \bibinfo{year}{2001}\natexlab{}.
\newblock \bibinfo{title}{Assessing Parenting Competence in Child Protection Cases: A Clinical Practice Model}.
\newblock
\newblock
Issue 1.


\bibitem[Clark(2012a)]%
        {Schofield2012}
\bibfield{author}{\bibinfo{person}{Lynn~Schofield Clark}.} \bibinfo{year}{2012}\natexlab{a}.
\newblock \showarticletitle{{75Identity 2.0: Young People and Digital and Mobile Media}}.
\newblock In \bibinfo{booktitle}{\emph{{The Parent App: Understanding Families in the Digital Age}}}. \bibinfo{publisher}{Oxford University Press}.
\newblock
\showISBNx{9780199899616}
\urldef\tempurl%
\url{https://doi.org/10.1093/acprof:oso/9780199899616.003.0004}
\showDOI{\tempurl}
\showeprint{https://academic.oup.com/book/0/chapter/150322210/chapter-ag-pdf/44982177/book\_6470\_section\_150322210.ag.pdf}


\bibitem[Clark(2012b)]%
        {ClarkBookChapter9}
\bibfield{author}{\bibinfo{person}{Lynn~Schofield Clark}.} \bibinfo{year}{2012}\natexlab{b}.
\newblock \showarticletitle{{Parenting in a Digital Age: The Mediatization of Family Life and the Need to Act}}.
\newblock In \bibinfo{booktitle}{\emph{{The Parent App: Understanding Families in the Digital Age}}}. \bibinfo{publisher}{Oxford University Press}, \bibinfo{pages}{201--226}.
\newblock
\showISBNx{9780199899616}
\urldef\tempurl%
\url{https://doi.org/10.1093/acprof:oso/9780199899616.003.0009}
\showDOI{\tempurl}
\showeprint{https://academic.oup.com/book/0/chapter/150328769/chapter-ag-pdf/44982173/book\_6470\_section\_150328769.ag.pdf}


\bibitem[Clark(2012c)]%
        {ClarkBookChapter1}
\bibfield{author}{\bibinfo{person}{Lynn~Schofield Clark}.} \bibinfo{year}{2012}\natexlab{c}.
\newblock \showarticletitle{{Risk, Media, and Parenting in a Digital Age}}.
\newblock In \bibinfo{booktitle}{\emph{{The Parent App: Understanding Families in the Digital Age}}}. \bibinfo{publisher}{Oxford University Press}, \bibinfo{pages}{3--27}.
\newblock
\showISBNx{9780199899616}
\urldef\tempurl%
\url{https://doi.org/10.1093/acprof:oso/9780199899616.003.0001}
\showDOI{\tempurl}


\bibitem[Clarke and Sulsky(2019)]%
        {Clarke2019}
\bibfield{author}{\bibinfo{person}{Heather~M. Clarke} {and} \bibinfo{person}{Lorne~M. Sulsky}.} \bibinfo{year}{2019}\natexlab{}.
\newblock \showarticletitle{The Impact of Gender Stereotypes on the Appraisal of Civic Virtue Performance}.
\newblock \bibinfo{journal}{\emph{Journal of Research in Gender Studies}}  \bibinfo{volume}{9} (\bibinfo{year}{2019}).
\newblock
\urldef\tempurl%
\url{https://heinonline.org/HOL/Page?handle=hein.journals/jogenst9&id=221&div=19&collection=journals}
\showURL{%
\tempurl}


\bibitem[Costello and Ramo(2017)]%
        {Costello2017}
\bibfield{author}{\bibinfo{person}{Caitlin~R. Costello} {and} \bibinfo{person}{Danielle~E. Ramo}.} \bibinfo{year}{2017}\natexlab{}.
\newblock \showarticletitle{Social Media and Substance Use: What Should We Be Recommending to Teens and Their Parents?}
\newblock \bibinfo{journal}{\emph{Journal of Adolescent Health}}  \bibinfo{volume}{60} (\bibinfo{date}{6} \bibinfo{year}{2017}), \bibinfo{pages}{629--630}.
\newblock
Issue 6.
\showISSN{18791972}
\urldef\tempurl%
\url{https://doi.org/10.1016/j.jadohealth.2017.03.017}
\showDOI{\tempurl}


\bibitem[Cranor et~al\mbox{.}(2014)]%
        {lorrie2014}
\bibfield{author}{\bibinfo{person}{Lorrie~Faith Cranor}, \bibinfo{person}{Adam~L Durity}, \bibinfo{person}{Abigail Marsh}, {and} \bibinfo{person}{Blase Ur}.} \bibinfo{year}{2014}\natexlab{}.
\newblock \showarticletitle{Parents' and Teens' Perspectives on Privacy In a Technology-Filled World}.
\newblock  (\bibinfo{year}{2014}).
\newblock


\bibitem[Danet(2020)]%
        {Danet2020}
\bibfield{author}{\bibinfo{person}{Marie Danet}.} \bibinfo{year}{2020}\natexlab{}.
\newblock \showarticletitle{Parental Concerns about their School-aged Children’s Use of Digital Devices}.
\newblock \bibinfo{journal}{\emph{Journal of Child and Family Studies}}  \bibinfo{volume}{29} (\bibinfo{date}{10} \bibinfo{year}{2020}).
\newblock
\urldef\tempurl%
\url{https://doi.org/10.1007/s10826-020-01760-y}
\showDOI{\tempurl}


\bibitem[Deldari et~al\mbox{.}(2023)]%
        {Deldari2023}
\bibfield{author}{\bibinfo{person}{Elmira Deldari}, \bibinfo{person}{Diana Freed}, \bibinfo{person}{Julio Poveda}, {and} \bibinfo{person}{Yaxing Yao}.} \bibinfo{year}{2023}\natexlab{}.
\newblock \bibinfo{booktitle}{\emph{An Investigation of Teenager Experiences in Social Virtual Reality from Teenagers', Parents', and Bystanders' Perspectives}}.
\newblock 1--17 pages.
\newblock
\showISBNx{978-1-939133-36-6}
\urldef\tempurl%
\url{https://www.usenix.org/conference/soups2023/presentation/deldari}
\showURL{%
\tempurl}


\bibitem[Didehbani et~al\mbox{.}(2016)]%
        {Didehbani2016}
\bibfield{author}{\bibinfo{person}{Nyaz Didehbani}, \bibinfo{person}{Tandra Allen}, \bibinfo{person}{Michelle Kandalaft}, \bibinfo{person}{Daniel Krawczyk}, {and} \bibinfo{person}{Sandra Chapman}.} \bibinfo{year}{2016}\natexlab{}.
\newblock \showarticletitle{Virtual Reality Social Cognition Training for children with high functioning autism}.
\newblock \bibinfo{journal}{\emph{Computers in Human Behavior}}  \bibinfo{volume}{62} (\bibinfo{date}{9} \bibinfo{year}{2016}), \bibinfo{pages}{703--711}.
\newblock
\showISSN{0747-5632}
\urldef\tempurl%
\url{https://doi.org/10.1016/J.CHB.2016.04.033}
\showDOI{\tempurl}


\bibitem[Elkin et~al\mbox{.}(2021)]%
        {Elkin2021}
\bibfield{author}{\bibinfo{person}{Lisa~A Elkin}, \bibinfo{person}{Paul~G Allen}, \bibinfo{person}{Matthew Kay}, \bibinfo{person}{James~J Higgins}, {and} \bibinfo{person}{Jacob~O Wobbrock}.} \bibinfo{year}{2021}\natexlab{}.
\newblock \showarticletitle{An Aligned Rank Transform Procedure for Multifactor Contrast Tests; An Aligned Rank Transform Procedure for Multifactor Contrast Tests}.
\newblock   \bibinfo{volume}{15} (\bibinfo{year}{2021}).
\newblock
Issue 21.
\showISBNx{9781450386357}
\urldef\tempurl%
\url{https://doi.org/10.1145/3472749.3474784}
\showDOI{\tempurl}


\bibitem[Erickson et~al\mbox{.}(2016)]%
        {Erickson2016}
\bibfield{author}{\bibinfo{person}{Lee~B. Erickson}, \bibinfo{person}{Pamela Wisniewski}, \bibinfo{person}{Heng Xu}, \bibinfo{person}{John~M. Carroll}, \bibinfo{person}{Mary~Beth Rosson}, {and} \bibinfo{person}{Daniel~F. Perkins}.} \bibinfo{year}{2016}\natexlab{}.
\newblock \showarticletitle{The boundaries between: Parental involvement in a teen's online world}.
\newblock \bibinfo{journal}{\emph{Journal of the Association for Information Science and Technology}} \bibinfo{volume}{67}, \bibinfo{number}{6} (\bibinfo{year}{2016}), \bibinfo{pages}{1384--1403}.
\newblock
\urldef\tempurl%
\url{https://doi.org/10.1002/asi.23450}
\showDOI{\tempurl}
\showeprint{https://asistdl.onlinelibrary.wiley.com/doi/pdf/10.1002/asi.23450}


\bibitem[Fiani et~al\mbox{.}(2023)]%
        {fiani2023IDC}
\bibfield{author}{\bibinfo{person}{Cristina Fiani}, \bibinfo{person}{Robin Bretin}, \bibinfo{person}{Mark Mcgill}, {and} \bibinfo{person}{Mohamed Khamis}.} \bibinfo{year}{2023}\natexlab{}.
\newblock \showarticletitle{Big Buddy: Exploring Child Reactions and Parental Perceptions towards a Simulated Embodied Moderating System for Social Virtual Reality}. In \bibinfo{booktitle}{\emph{Proceedings of the 22nd Annual ACM Interaction Design and Children Conference}} (Chicago, IL, USA) \emph{(\bibinfo{series}{IDC '23})}. \bibinfo{publisher}{Association for Computing Machinery}, \bibinfo{address}{New York, NY, USA}, \bibinfo{pages}{1–13}.
\newblock
\showISBNx{9798400701313}
\urldef\tempurl%
\url{https://doi.org/10.1145/3585088.3589374}
\showDOI{\tempurl}


\bibitem[Flanagan(1954)]%
        {Flanagan1954}
\bibfield{author}{\bibinfo{person}{John~C. Flanagan}.} \bibinfo{year}{1954}\natexlab{}.
\newblock \showarticletitle{The critical incident technique}.
\newblock \bibinfo{journal}{\emph{Psychological Bulletin}}  \bibinfo{volume}{51} (\bibinfo{date}{7} \bibinfo{year}{1954}), \bibinfo{pages}{327--358}.
\newblock
Issue 4.
\showISSN{00332909}
\urldef\tempurl%
\url{https://doi.org/10.1037/H0061470}
\showDOI{\tempurl}


\bibitem[Freeman et~al\mbox{.}(2022)]%
        {Freeman2022}
\bibfield{author}{\bibinfo{person}{Guo Freeman}, \bibinfo{person}{Samaneh Zamanifard}, \bibinfo{person}{Divine Maloney}, {and} \bibinfo{person}{Dane Acena}.} \bibinfo{year}{2022}\natexlab{}.
\newblock \showarticletitle{Disturbing the Peace: Experiencing and Mitigating Emerging Harassment in Social Virtual Reality}.
\newblock \bibinfo{journal}{\emph{Proc. ACM Hum.-Comput. Interact.}} \bibinfo{volume}{6}, \bibinfo{number}{CSCW1}, Article \bibinfo{articleno}{85} (\bibinfo{date}{apr} \bibinfo{year}{2022}), \bibinfo{numpages}{30}~pages.
\newblock
\urldef\tempurl%
\url{https://doi.org/10.1145/3512932}
\showDOI{\tempurl}


\bibitem[Gaggioli(2018)]%
        {Gaggioli2018}
\bibfield{author}{\bibinfo{person}{Andrea Gaggioli}.} \bibinfo{year}{2018}\natexlab{}.
\newblock \showarticletitle{Virtually Social}.
\newblock \bibinfo{journal}{\emph{CYBERPSYCHOLOGY, BEHAVIOR AND SOCIAL NETWORKING}}  \bibinfo{volume}{21} (\bibinfo{date}{5} \bibinfo{year}{2018}), \bibinfo{pages}{338--339}.
\newblock
Issue 5.
\showISSN{2152-2715}
\urldef\tempurl%
\url{https://doi.org/10.1089/CYBER.2018.29112.CSI}
\showDOI{\tempurl}


\bibitem[Glassner(1999)]%
        {Glassner1999}
\bibfield{author}{\bibinfo{person}{Barry Glassner}.} \bibinfo{year}{1999}\natexlab{}.
\newblock \bibinfo{booktitle}{\emph{The culture of fear: Why Americans are afraid of the wrong things.}}
\newblock \bibinfo{publisher}{Basic Books}. xix--xxviii pages.
\newblock


\bibitem[Greenfield(2004)]%
        {GREENFIELD2004751}
\bibfield{author}{\bibinfo{person}{Patricia~M. Greenfield}.} \bibinfo{year}{2004}\natexlab{}.
\newblock \showarticletitle{Developmental considerations for determining appropriate Internet use guidelines for children and adolescents}.
\newblock \bibinfo{journal}{\emph{Journal of Applied Developmental Psychology}} \bibinfo{volume}{25}, \bibinfo{number}{6} (\bibinfo{year}{2004}), \bibinfo{pages}{751--762}.
\newblock
\showISSN{0193-3973}
\urldef\tempurl%
\url{https://doi.org/10.1016/j.appdev.2004.09.008}
\showDOI{\tempurl}
\newblock
\shownote{Developing Children, Developing Media - Research from Television to the Internet from the Children's Digital Media Center: A Special Issue Dedicated to the Memory of Rodney R. Cocking}.


\bibitem[Hartikainen et~al\mbox{.}(2016)]%
        {Hartikainen2016}
\bibfield{author}{\bibinfo{person}{Heidi Hartikainen}, \bibinfo{person}{Netta Iivari}, {and} \bibinfo{person}{Marianne Kinnula}.} \bibinfo{year}{2016}\natexlab{}.
\newblock \showarticletitle{Should We design for control, trust or involvement? A discourses survey about children's online safety}.
\newblock \bibinfo{journal}{\emph{Proceedings of IDC 2016 - The 15th International Conference on Interaction Design and Children}} (\bibinfo{date}{6} \bibinfo{year}{2016}), \bibinfo{pages}{367--378}.
\newblock
\showISBNx{9781450343138}
\urldef\tempurl%
\url{https://doi.org/10.1145/2930674.2930680}
\showDOI{\tempurl}


\bibitem[Hiniker et~al\mbox{.}(2016)]%
        {Hiniker2016}
\bibfield{author}{\bibinfo{person}{Alexis Hiniker}, \bibinfo{person}{Sarita~Y. Schoenebeck}, {and} \bibinfo{person}{Julie~A. Kientz}.} \bibinfo{year}{2016}\natexlab{}.
\newblock \showarticletitle{Not at the dinner table: Parents' and children's perspectives on family technology rules}.
\newblock \bibinfo{journal}{\emph{Proceedings of the ACM Conference on Computer Supported Cooperative Work, CSCW}}  \bibinfo{volume}{27} (\bibinfo{date}{2} \bibinfo{year}{2016}), \bibinfo{pages}{1376--1389}.
\newblock
\showISBNx{9781450335928}
\urldef\tempurl%
\url{https://doi.org/10.1145/2818048.2819940}
\showDOI{\tempurl}


\bibitem[Huber et~al\mbox{.}(2018)]%
        {Huber2018}
\bibfield{author}{\bibinfo{person}{Brittany Huber}, \bibinfo{person}{Kate Highfield}, {and} \bibinfo{person}{Jordy Kaufman}.} \bibinfo{year}{2018}\natexlab{}.
\newblock \showarticletitle{Detailing the digital experience: Parent reports of children's media use in the home learning environment}.
\newblock \bibinfo{journal}{\emph{British Journal of Educational Technology}}  \bibinfo{volume}{49} (\bibinfo{date}{9} \bibinfo{year}{2018}), \bibinfo{pages}{821--833}.
\newblock
Issue 5.
\showISSN{1467-8535}
\urldef\tempurl%
\url{https://doi.org/10.1111/BJET.12667}
\showDOI{\tempurl}


\bibitem[Ito et~al\mbox{.}(2018)]%
        {Ito2018}
\bibfield{author}{\bibinfo{person}{Mizuko Ito}, \bibinfo{person}{Sonja Baumer}, \bibinfo{person}{Matteo Bittanti}, \bibinfo{person}{danah boyd}, \bibinfo{person}{Rachel Cody}, \bibinfo{person}{Becky~Herr Stephenson}, \bibinfo{person}{Heather~A. Horst}, \bibinfo{person}{Patricia~G. Lange}, \bibinfo{person}{Dilan Mahendran}, \bibinfo{person}{Katynka~Z. Martínez}, \bibinfo{person}{C.~J. Pascoe}, \bibinfo{person}{Dan Perkel}, \bibinfo{person}{Laura Robinson}, \bibinfo{person}{Christo Sims}, {and} \bibinfo{person}{Lisa Tripp}.} \bibinfo{year}{2018}\natexlab{}.
\newblock \showarticletitle{Gaming}.
\newblock In \bibinfo{booktitle}{\emph{Hanging Out, Messing Around, and Geeking Out}}. \bibinfo{publisher}{The MIT Press}, Chapter~5.
\newblock
\urldef\tempurl%
\url{https://doi.org/10.7551/MITPRESS/8402.001.0001}
\showDOI{\tempurl}


\bibitem[Kaimara et~al\mbox{.}(2021)]%
        {Kaimara2021}
\bibfield{author}{\bibinfo{person}{Polyxeni Kaimara}, \bibinfo{person}{Andreas Oikonomou}, {and} \bibinfo{person}{Ioannis Deliyannis}.} \bibinfo{year}{2021}\natexlab{}.
\newblock \showarticletitle{Could virtual reality applications pose real risks to children and adolescents? A systematic review of ethical issues and concerns}.
\newblock \bibinfo{journal}{\emph{Virtual Reality 2021 26:2}}  \bibinfo{volume}{26} (\bibinfo{date}{8} \bibinfo{year}{2021}), \bibinfo{pages}{697--735}.
\newblock
Issue 2.
\showISBNx{0123456789}
\showISSN{1434-9957}
\urldef\tempurl%
\url{https://doi.org/10.1007/S10055-021-00563-W}
\showDOI{\tempurl}


\bibitem[Mado et~al\mbox{.}(2022)]%
        {Mado2022}
\bibfield{author}{\bibinfo{person}{Marijn Mado}, \bibinfo{person}{Géraldine Fauville}, \bibinfo{person}{Hanseul Jun}, \bibinfo{person}{Elise Most}, \bibinfo{person}{Carlyn Strang}, {and} \bibinfo{person}{Jeremy~N. Bailenson}.} \bibinfo{year}{2022}\natexlab{}.
\newblock \showarticletitle{Accessibility of Educational Virtual Reality for Children During the COVID-19 Pandemic}.
\newblock \bibinfo{journal}{\emph{Technology, Mind, and Behavior}}  \bibinfo{volume}{3} (\bibinfo{date}{3} \bibinfo{year}{2022}).
\newblock
Issue 1.
\urldef\tempurl%
\url{https://doi.org/10.1037/TMB0000066}
\showDOI{\tempurl}


\bibitem[Maloney et~al\mbox{.}(2020a)]%
        {Maloney2020b}
\bibfield{author}{\bibinfo{person}{Divine Maloney}, \bibinfo{person}{Guo Freeman}, {and} \bibinfo{person}{Andrew Robb}.} \bibinfo{year}{2020}\natexlab{a}.
\newblock \showarticletitle{It Is Complicated: Interacting with Children in Social Virtual Reality}.
\newblock \bibinfo{journal}{\emph{Proceedings - 2020 IEEE Conference on Virtual Reality and 3D User Interfaces, VRW 2020}} (\bibinfo{date}{3} \bibinfo{year}{2020}), \bibinfo{pages}{343--347}.
\newblock
\showISBNx{9781728165325}
\urldef\tempurl%
\url{https://doi.org/10.1109/VRW50115.2020.00075}
\showDOI{\tempurl}


\bibitem[Maloney et~al\mbox{.}(2020b)]%
        {Maloney2020a}
\bibfield{author}{\bibinfo{person}{Divine Maloney}, \bibinfo{person}{Guo Freeman}, {and} \bibinfo{person}{Andrew Robb}.} \bibinfo{year}{2020}\natexlab{b}.
\newblock \showarticletitle{A Virtual Space for All: Exploring Children's Experience in Social Virtual Reality}.
\newblock \bibinfo{journal}{\emph{CHI PLAY 2020 - Proceedings of the Annual Symposium on Computer-Human Interaction in Play}}, \bibinfo{pages}{472--483}.
\newblock
\showISBNx{9781450380744}
\urldef\tempurl%
\url{https://doi.org/10.1145/3410404.3414268}
\showDOI{\tempurl}


\bibitem[Maloney et~al\mbox{.}(2021)]%
        {Maloney2021}
\bibfield{author}{\bibinfo{person}{Divine Maloney}, \bibinfo{person}{Guo Freeman}, {and} \bibinfo{person}{Andrew Robb}.} \bibinfo{year}{2021}\natexlab{}.
\newblock \showarticletitle{Stay Connected in An Immersive World: Why Teenagers Engage in Social Virtual Reality}.
\newblock \bibinfo{journal}{\emph{Proceedings of Interaction Design and Children, IDC 2021}}, \bibinfo{pages}{69--79}.
\newblock
\showISBNx{9781450384520}
\urldef\tempurl%
\url{https://doi.org/10.1145/3459990.3460703}
\showDOI{\tempurl}


\bibitem[McVeigh-Schultz et~al\mbox{.}(2018)]%
        {10.1145/3197391.3205451}
\bibfield{author}{\bibinfo{person}{Joshua McVeigh-Schultz}, \bibinfo{person}{Elena M\'{a}rquez~Segura}, \bibinfo{person}{Nick Merrill}, {and} \bibinfo{person}{Katherine Isbister}.} \bibinfo{year}{2018}\natexlab{}.
\newblock \showarticletitle{What's It Mean to "Be Social" in VR? Mapping the Social VR Design Ecology}. In \bibinfo{booktitle}{\emph{Proceedings of the 2018 ACM Conference Companion Publication on Designing Interactive Systems}} (Hong Kong, China) \emph{(\bibinfo{series}{DIS '18 Companion})}. \bibinfo{publisher}{Association for Computing Machinery}, \bibinfo{address}{New York, NY, USA}, \bibinfo{pages}{289–294}.
\newblock
\showISBNx{9781450356312}
\urldef\tempurl%
\url{https://doi.org/10.1145/3197391.3205451}
\showDOI{\tempurl}


\bibitem[Miehlbradt et~al\mbox{.}(2021)]%
        {Miehlbradt2021}
\bibfield{author}{\bibinfo{person}{Jenifer Miehlbradt}, \bibinfo{person}{Luigi~F. Cuturi}, \bibinfo{person}{Silvia Zanchi}, \bibinfo{person}{Monica Gori}, {and} \bibinfo{person}{Silvestro Micera}.} \bibinfo{year}{2021}\natexlab{}.
\newblock \showarticletitle{Immersive virtual reality interferes with default head–trunk coordination strategies in young children}.
\newblock \bibinfo{journal}{\emph{Scientific Reports}}  \bibinfo{volume}{11} (\bibinfo{date}{12} \bibinfo{year}{2021}), \bibinfo{pages}{17959}.
\newblock
Issue 1.
\showISSN{20452322}
\urldef\tempurl%
\url{https://doi.org/10.1038/s41598-021-96866-8}
\showDOI{\tempurl}


\bibitem[Morris and Fritz(2013)]%
        {Morris2013}
\bibfield{author}{\bibinfo{person}{Peter~E. Morris} {and} \bibinfo{person}{Catherine~O. Fritz}.} \bibinfo{year}{2013}\natexlab{}.
\newblock \showarticletitle{Effect sizes in memory research}.
\newblock \bibinfo{journal}{\emph{http://dx.doi.org/10.1080/09658211.2013.763984}}  \bibinfo{volume}{21} (\bibinfo{date}{10} \bibinfo{year}{2013}), \bibinfo{pages}{832--842}.
\newblock
Issue 7.
\showISSN{09658211}
\urldef\tempurl%
\url{https://doi.org/10.1080/09658211.2013.763984}
\showDOI{\tempurl}


\bibitem[Nowell et~al\mbox{.}(2017)]%
        {Nowell2017}
\bibfield{author}{\bibinfo{person}{Lorelli~S. Nowell}, \bibinfo{person}{Jill~M. Norris}, \bibinfo{person}{Deborah~E. White}, {and} \bibinfo{person}{Nancy~J. Moules}.} \bibinfo{year}{2017}\natexlab{}.
\newblock \showarticletitle{Thematic Analysis: Striving to Meet the Trustworthiness Criteria}.
\newblock \bibinfo{journal}{\emph{https://doi.org/10.1177/1609406917733847}}  \bibinfo{volume}{16} (\bibinfo{date}{10} \bibinfo{year}{2017}).
\newblock
Issue 1.
\showISSN{16094069}
\urldef\tempurl%
\url{https://doi.org/10.1177/1609406917733847}
\showDOI{\tempurl}


\bibitem[Odgers et~al\mbox{.}(2020)]%
        {Odgers}
\bibfield{author}{\bibinfo{person}{Candice~L. Odgers}, \bibinfo{person}{Stephen~M. Schueller}, {and} \bibinfo{person}{Mimi Ito}.} \bibinfo{year}{2020}\natexlab{}.
\newblock \showarticletitle{Screen Time, Social Media Use, and Adolescent Development}.
\newblock \bibinfo{journal}{\emph{Annual Review of Developmental Psychology}} \bibinfo{volume}{2}, \bibinfo{number}{1} (\bibinfo{year}{2020}), \bibinfo{pages}{485--502}.
\newblock
\urldef\tempurl%
\url{https://doi.org/10.1146/annurev-devpsych-121318-084815}
\showDOI{\tempurl}


\bibitem[O’Hagan et~al\mbox{.}(2021)]%
        {joseph2021}
\bibfield{author}{\bibinfo{person}{Joseph O’Hagan}, \bibinfo{person}{Julie~R. Williamson}, \bibinfo{person}{Mark McGill}, {and} \bibinfo{person}{Mohamed Khamis}.} \bibinfo{year}{2021}\natexlab{}.
\newblock \showarticletitle{Safety, Power Imbalances, Ethics and Proxy Sex: Surveying In-The-Wild Interactions Between VR Users and Bystanders}. In \bibinfo{booktitle}{\emph{2021 IEEE International Symposium on Mixed and Augmented Reality (ISMAR)}}. \bibinfo{pages}{211--220}.
\newblock
\urldef\tempurl%
\url{https://doi.org/10.1109/ISMAR52148.2021.00036}
\showDOI{\tempurl}


\bibitem[Pasquale et~al\mbox{.}(2022)]%
        {Pasquale2020}
\bibfield{author}{\bibinfo{person}{Liliana Pasquale}, \bibinfo{person}{Paola Zippo}, \bibinfo{person}{Cliona Curley}, \bibinfo{person}{Brian O’Neill}, {and} \bibinfo{person}{Marina Mongiello}.} \bibinfo{year}{2022}\natexlab{}.
\newblock \showarticletitle{Digital Age of Consent and Age Verification: Can They Protect Children?}
\newblock \bibinfo{journal}{\emph{IEEE Software}} \bibinfo{volume}{39}, \bibinfo{number}{3} (\bibinfo{year}{2022}), \bibinfo{pages}{50--57}.
\newblock
\urldef\tempurl%
\url{https://doi.org/10.1109/MS.2020.3044872}
\showDOI{\tempurl}


\bibitem[Ph.D. and Ph.D.(2008)]%
        {Livinstone2008}
\bibfield{author}{\bibinfo{person}{Sonia~Livingstone Ph.D.} {and} \bibinfo{person}{Ellen J.~Helsper Ph.D.}} \bibinfo{year}{2008}\natexlab{}.
\newblock \showarticletitle{Parental Mediation of Children's Internet Use}.
\newblock \bibinfo{journal}{\emph{Journal of Broadcasting \& Electronic Media}} \bibinfo{volume}{52}, \bibinfo{number}{4} (\bibinfo{year}{2008}), \bibinfo{pages}{581--599}.
\newblock
\urldef\tempurl%
\url{https://doi.org/10.1080/08838150802437396}
\showDOI{\tempurl}


\bibitem[Phippen(2022)]%
        {Phippen2022}
\bibfield{author}{\bibinfo{person}{Andy Phippen}.} \bibinfo{year}{2022}\natexlab{}.
\newblock \showarticletitle{Protecting children in the metaverse: it’s easy to blame big tech, but we all have a role to play}.
\newblock  (\bibinfo{date}{3} \bibinfo{year}{2022}).
\newblock
\urldef\tempurl%
\url{https://blogs.lse.ac.uk/parenting4digitalfuture/}
\showURL{%
\tempurl}
\newblock
\shownote{Last Accessed: 02-10-2023}.


\bibitem[Radesky et~al\mbox{.}(2016)]%
        {Radesky2016}
\bibfield{author}{\bibinfo{person}{Jenny~S. Radesky}, \bibinfo{person}{Caroline Kistin}, \bibinfo{person}{Staci Eisenberg}, \bibinfo{person}{Jamie Gross}, \bibinfo{person}{Gabrielle Block}, \bibinfo{person}{Barry Zuckerman}, {and} \bibinfo{person}{Michael Silverstein}.} \bibinfo{year}{2016}\natexlab{}.
\newblock \showarticletitle{Parent Perspectives on Their Mobile Technology Use: The Excitement and Exhaustion of Parenting while Connected}.
\newblock \bibinfo{journal}{\emph{Journal of Developmental and Behavioral Pediatrics}}  \bibinfo{volume}{37} (\bibinfo{year}{2016}), \bibinfo{pages}{694--701}.
\newblock
Issue 9.
\showISSN{15367312}
\urldef\tempurl%
\url{https://doi.org/10.1097/DBP.0000000000000357}
\showDOI{\tempurl}


\bibitem[Richardson(2011)]%
        {RICHARDSON2011135}
\bibfield{author}{\bibinfo{person}{John~T.E. Richardson}.} \bibinfo{year}{2011}\natexlab{}.
\newblock \showarticletitle{Eta squared and partial eta squared as measures of effect size in educational research}.
\newblock \bibinfo{journal}{\emph{Educational Research Review}} \bibinfo{volume}{6}, \bibinfo{number}{2} (\bibinfo{year}{2011}), \bibinfo{pages}{135--147}.
\newblock
\showISSN{1747-938X}
\urldef\tempurl%
\url{https://doi.org/10.1016/j.edurev.2010.12.001}
\showDOI{\tempurl}


\bibitem[Russell(1980)]%
        {Russell1980}
\bibfield{author}{\bibinfo{person}{James~A. Russell}.} \bibinfo{year}{1980}\natexlab{}.
\newblock \showarticletitle{A circumplex model of affect}.
\newblock \bibinfo{journal}{\emph{Journal of Personality and Social Psychology}}  \bibinfo{volume}{39} (\bibinfo{date}{12} \bibinfo{year}{1980}), \bibinfo{pages}{1161--1178}.
\newblock
Issue 6.
\showISSN{00223514}
\urldef\tempurl%
\url{https://doi.org/10.1037/H0077714}
\showDOI{\tempurl}


\bibitem[Sameer(nd)]%
        {blogpost}
\bibfield{author}{\bibinfo{person}{Hinduja Sameer}.} \bibinfo{year}{n.d}\natexlab{}.
\newblock \bibinfo{title}{Child Grooming and the Metaverse – Issues and Solutions - Cyberbullying Research Center}.
\newblock
\newblock
\urldef\tempurl%
\url{https://cyberbullying.org/child-grooming-metaverse}
\showURL{%
\tempurl}
\newblock
\shownote{Last Accessed: 02-10-2023}.


\bibitem[Schulenberg et~al\mbox{.}(2023)]%
        {schulenberg2023towards}
\bibfield{author}{\bibinfo{person}{Kelsea Schulenberg}, \bibinfo{person}{Lingyuan Li}, \bibinfo{person}{Guo Freeman}, \bibinfo{person}{Samaneh Zamanifard}, {and} \bibinfo{person}{Nathan~J. McNeese}.} \bibinfo{year}{2023}\natexlab{}.
\newblock \showarticletitle{Towards Leveraging AI-based Moderation to Address Emergent Harassment in Social Virtual Reality}.
\newblock  (\bibinfo{year}{2023}), \bibinfo{pages}{17}.
\newblock
\urldef\tempurl%
\url{https://doi.org/10.1145/3544548.3581090}
\showDOI{\tempurl}


\bibitem[Sciacca et~al\mbox{.}(2022)]%
        {Sciacca2022}
\bibfield{author}{\bibinfo{person}{Beatrice Sciacca}, \bibinfo{person}{Derek~A. Laffan}, \bibinfo{person}{James~O'Higgins Norman}, {and} \bibinfo{person}{Tijana Milosevic}.} \bibinfo{year}{2022}\natexlab{}.
\newblock \showarticletitle{Parental mediation in pandemic: Predictors and relationship with children's digital skills and time spent online in Ireland}.
\newblock \bibinfo{journal}{\emph{Computers in Human Behavior}}  \bibinfo{volume}{127} (\bibinfo{date}{2} \bibinfo{year}{2022}), \bibinfo{pages}{107081}.
\newblock
\showISSN{07475632}
\urldef\tempurl%
\url{https://doi.org/10.1016/j.chb.2021.107081}
\showDOI{\tempurl}


\bibitem[Slupska and Tanczer(2021)]%
        {Slupska2021}
\bibfield{author}{\bibinfo{person}{Julia Slupska} {and} \bibinfo{person}{Leonie~Maria Tanczer}.} \bibinfo{year}{2021}\natexlab{}.
\newblock \showarticletitle{Threat Modeling Intimate Partner Violence: Tech Abuse as a Cybersecurity Challenge in the Internet of Things}.
\newblock \bibinfo{journal}{\emph{The Emerald International Handbook of Technology-Facilitated Violence and Abuse}} (\bibinfo{date}{6} \bibinfo{year}{2021}), \bibinfo{pages}{663--688}.
\newblock
\showISBNx{978-1-83982-849-2}
\urldef\tempurl%
\url{https://doi.org/10.1108/978-1-83982-848-520211049}
\showDOI{\tempurl}


\bibitem[Smahelova et~al\mbox{.}(2017)]%
        {Smahelova2017}
\bibfield{author}{\bibinfo{person}{Martina Smahelova}, \bibinfo{person}{Dana Juhová}, \bibinfo{person}{Ivo Cermak}, {and} \bibinfo{person}{David Smahel}.} \bibinfo{year}{2017}\natexlab{}.
\newblock \showarticletitle{Mediation of young children’s digital technology use: The parents’ perspective}.
\newblock \bibinfo{journal}{\emph{Cyberpsychology: Journal of Psychosocial Research on Cyberspace}} \bibinfo{volume}{11}, \bibinfo{number}{3} (\bibinfo{date}{Nov.} \bibinfo{year}{2017}), \bibinfo{pages}{Article 4}.
\newblock
\urldef\tempurl%
\url{https://doi.org/10.5817/CP2017-3-4}
\showDOI{\tempurl}


\bibitem[Steinkuehler(2016)]%
        {Steinkuehler2016}
\bibfield{author}{\bibinfo{person}{Constance Steinkuehler}.} \bibinfo{year}{2016}\natexlab{}.
\newblock \showarticletitle{Parenting and Video Games}.
\newblock \bibinfo{journal}{\emph{Journal of Adolescent \& Adult Literacy}} \bibinfo{volume}{59}, \bibinfo{number}{4} (\bibinfo{year}{2016}), \bibinfo{pages}{357--361}.
\newblock
\urldef\tempurl%
\url{https://doi.org/10.1002/jaal.455}
\showDOI{\tempurl}
\showeprint{https://ila.onlinelibrary.wiley.com/doi/pdf/10.1002/jaal.455}


\bibitem[Sykownik et~al\mbox{.}(2022)]%
        {Sykownik2022}
\bibfield{author}{\bibinfo{person}{Philipp Sykownik}, \bibinfo{person}{Divine Maloney}, \bibinfo{person}{Guo Freeman}, {and} \bibinfo{person}{Maic Masuch}.} \bibinfo{year}{2022}\natexlab{}.
\newblock \showarticletitle{Something Personal from the Metaverse: Goals, Topics, and Contextual Factors of Self-Disclosure in Commercial Social VR}. In \bibinfo{booktitle}{\emph{Proceedings of the 2022 CHI Conference on Human Factors in Computing Systems}} (New Orleans, LA, USA) \emph{(\bibinfo{series}{CHI '22})}. \bibinfo{publisher}{Association for Computing Machinery}, \bibinfo{address}{New York, NY, USA}, Article \bibinfo{articleno}{632}, \bibinfo{numpages}{17}~pages.
\newblock
\showISBNx{9781450391573}
\urldef\tempurl%
\url{https://doi.org/10.1145/3491102.3502008}
\showDOI{\tempurl}


\bibitem[Tanczer et~al\mbox{.}(2021)]%
        {Tanczer2021}
\bibfield{author}{\bibinfo{person}{Leonie~Maria Tanczer}, \bibinfo{person}{Isabel López-Neira}, {and} \bibinfo{person}{Simon Parkin}.} \bibinfo{year}{2021}\natexlab{}.
\newblock \showarticletitle{‘I feel like we’re really behind the game’: perspectives of the United Kingdom’s intimate partner violence support sector on the rise of technology-facilitated abuse}.
\newblock \bibinfo{journal}{\emph{Journal of Gender-Based Violence}}  \bibinfo{volume}{5} (\bibinfo{date}{10} \bibinfo{year}{2021}), \bibinfo{pages}{431--450}.
\newblock
Issue 3.
\showISSN{2398-6808}
\urldef\tempurl%
\url{https://doi.org/10.1332/239868021X16290304343529}
\showDOI{\tempurl}


\bibitem[Tseng et~al\mbox{.}(2022)]%
        {tseng2022chi}
\bibfield{author}{\bibinfo{person}{Wen-Jie Tseng}, \bibinfo{person}{Elise Bonnail}, \bibinfo{person}{Mark McGill}, \bibinfo{person}{Mohamed Khamis}, \bibinfo{person}{Eric Lecolinet}, \bibinfo{person}{Samuel Huron}, {and} \bibinfo{person}{Jan Gugenheimer}.} \bibinfo{year}{2022}\natexlab{}.
\newblock \showarticletitle{The Dark Side of Perceptual Manipulations in Virtual Reality}. In \bibinfo{booktitle}{\emph{Proceedings of the 2022 CHI Conference on Human Factors in Computing Systems}} (New Orleans, LA, USA) \emph{(\bibinfo{series}{CHI '22})}. \bibinfo{publisher}{Association for Computing Machinery}, \bibinfo{address}{New York, NY, USA}, Article \bibinfo{articleno}{612}, \bibinfo{numpages}{15}~pages.
\newblock
\showISBNx{9781450391573}
\urldef\tempurl%
\url{https://doi.org/10.1145/3491102.3517728}
\showDOI{\tempurl}


\bibitem[Valkenburg et~al\mbox{.}(2022)]%
        {Valkenburg2022}
\bibfield{author}{\bibinfo{person}{Patti~M. Valkenburg}, \bibinfo{person}{Adrian Meier}, {and} \bibinfo{person}{Ine Beyens}.} \bibinfo{year}{2022}\natexlab{}.
\newblock \showarticletitle{Social media use and its impact on adolescent mental health: An umbrella review of the evidence}.
\newblock \bibinfo{journal}{\emph{Current Opinion in Psychology}}  \bibinfo{volume}{44} (\bibinfo{date}{4} \bibinfo{year}{2022}), \bibinfo{pages}{58--68}.
\newblock
\showISSN{2352250X}
\urldef\tempurl%
\url{https://doi.org/10.1016/j.copsyc.2021.08.017}
\showDOI{\tempurl}


\bibitem[Wilson and McGill(2018)]%
        {Wilson2018}
\bibfield{author}{\bibinfo{person}{Graham Wilson} {and} \bibinfo{person}{Mark McGill}.} \bibinfo{year}{2018}\natexlab{}.
\newblock \showarticletitle{Violent video games in virtual reality: Re-evaluating the impact and rating of interactive experiences}.
\newblock \bibinfo{journal}{\emph{CHI PLAY 2018 - Proceedings of the 2018 Annual Symposium on Computer-Human Interaction in Play}} (\bibinfo{date}{10} \bibinfo{year}{2018}), \bibinfo{pages}{73--85}.
\newblock
\showISBNx{9781450356244}
\urldef\tempurl%
\url{https://doi.org/10.1145/3242671.3242684}
\showDOI{\tempurl}


\bibitem[Wisniewski et~al\mbox{.}(2014)]%
        {Wisniewski2014}
\bibfield{author}{\bibinfo{person}{Pamela Wisniewski}, \bibinfo{person}{Haiyan Jia}, \bibinfo{person}{Heng Xu}, \bibinfo{person}{Mary~Beth Rosson}, {and} \bibinfo{person}{John Carroll}.} \bibinfo{year}{2014}\natexlab{}.
\newblock \showarticletitle{“Preventative” vs.“Reactive:” How Parental Mediation Influences Teens’ Social Media Privacy Behaviors}.
\newblock  (\bibinfo{date}{01} \bibinfo{year}{2014}).
\newblock


\bibitem[Zhang et~al\mbox{.}(2022)]%
        {Zhang2022}
\bibfield{author}{\bibinfo{person}{Minyue Zhang}, \bibinfo{person}{Hongwei Ding}, \bibinfo{person}{Meri Naumceska}, {and} \bibinfo{person}{Yang Zhang}.} \bibinfo{year}{2022}\natexlab{}.
\newblock \showarticletitle{Virtual Reality Technology as an Educational and Intervention Tool for Children with Autism Spectrum Disorder: Current Perspectives and Future Directions}.
\newblock \bibinfo{journal}{\emph{Behavioral Sciences 2022, Vol. 12, Page 138}}  \bibinfo{volume}{12} (\bibinfo{date}{5} \bibinfo{year}{2022}), \bibinfo{pages}{138}.
\newblock
Issue 5.
\showISSN{2076-328X}
\urldef\tempurl%
\url{https://doi.org/10.3390/BS12050138}
\showDOI{\tempurl}


\bibitem[Çankaya and Odabaşi(2009)]%
        {Serkan2009}
\bibfield{author}{\bibinfo{person}{Serkan Çankaya} {and} \bibinfo{person}{Hatice~Ferhan Odabaşi}.} \bibinfo{year}{2009}\natexlab{}.
\newblock \showarticletitle{Parental controls on children's computer and Internet use}.
\newblock \bibinfo{journal}{\emph{Procedia - Social and Behavioral Sciences}}  \bibinfo{volume}{1} (\bibinfo{date}{1} \bibinfo{year}{2009}), \bibinfo{pages}{1105--1109}.
\newblock
Issue 1.
\showISSN{1877-0428}
\urldef\tempurl%
\url{https://doi.org/10.1016/J.SBSPRO.2009.01.199}
\showDOI{\tempurl}


\end{thebibliography}



\end{document}